\documentclass[useAMS,usenatbib]{mn2e}          

\usepackage{graphicx}
\usepackage{times}
\usepackage[utf8x]{inputenc}
\usepackage{amsmath}
\usepackage{amssymb}
\usepackage{amsfonts}
\usepackage{array}
\usepackage{natbib}
\def\Ib{{\cal I}_b}
\def\Io{{\cal I}_o}
\def\Ii{{\cal I}_i}
\def\Ic{{\cal I}_c}

\def\faq{{f_1(q)}}
\def\fal{{f_1(\lambda)}}

\def\fbl{{f_2(\lambda)}}
\def\etal{\it et al.\ }

\def\aap{{\it Astronomy \& Astrophysics}}
\def\araa{{\it Annual Review of Astronomy \& Astrophysics}}
\def\aj{{\it Astronomycal Journal}}
\def\apj{{\it Astrophysical Journal}}
\def\apjl{{\it Astrophysical Journal Letters}}
\def\nat{{\it Nature}}
\def\mnras{{\it Monthly Notices of the Royal Astronomical Society}}
\def\apjs{{\it Astrophysical Journal Supplement Series}}
\title[Habitable Zones with Stable Orbits for Planets around Binary Systems]{Habitable Zones with Stable Orbits for Planets around Binary Systems}     %Header title goes on top of even pages

\author[L. G. Jaime L. Aguilar B. Pichardo]{Luisa G. Jaime$^{1} \thanks{E-mail: luisa@nucleares.unam.mx (LGJ)}$, Luis Aguilar$^{2}$, 
 and Barbara Pichardo$^{1}$  \\ $^{1}$Instituto de Astronom\'\i a,
  Universidad Nacional Aut\'onoma de M\'exico, Apdo. postal 70-264,
  Ciudad Universitaria, M\'exico \\ $^{2}$Instituto de
  Astronom\'ia, Universidad Nacional Aut\'onoma de M\'exico,
  Apdo. postal 877, 22800 Ensenada, M\'exico}

\begin{document}

\date{Accepted. Received ; in original form }
\pagerange{\pageref{firstpage}--\pageref{lastpage}} \pubyear{----}
\maketitle

\label{firstpage}

\begin{abstract}

A general formulation to compute habitable zones around binary stars
is presented.  A habitable zone in this context must satisfy two
separate conditions: a radiative one and one of dynamical
stability. For the case of single stars, the usual concept of
circumstellar habitable zone is based on the radiative condition only,
as the dynamical stability condition is taken for granted (assuming
minimal perturbation from other planets). For the radiative condition,
we extend the simple formulation of the circumstellar habitable zone
for single stars, to the case of eccentric stellar binary systems,
where two sources of luminosity at different orbital phases contribute
to the irradiance of their planetary circumstellar and circumbinary
regions. Our approach considers binaries with eccentric orbits and
guarantees that orbits in the computed habitable zone remain within it
at all orbital phases. For the dynamical stability condition, we use
the approach of invariant loops developed by Pichardo et al. (2005) to
find regions of stable, non-intersecting orbits, which is a robust
method to find stable regions in binary stars, as it is based in the
existence of integrals of motion. We apply the combined criteria to
calculate habitable zones for 64 binary stars in the solar
neighborhood with known orbital parameters, including some with
discovered planets. Formulae and interpolating tables are provided, so
the reader can compute the boundaries of the habitable zones for an
arbitrary binary system, using the stellar flux limits they
prefer. Together with the formulae provided for stable zones, these
allow the computation of both regions of stability and habitability
around any binary stellar system. We found 56\% of the cases we
consider can satisfy both restrictions, this is a very important
constriction to binary systems. Nevertheless, we conclude that these
systems where a dynamical and radiative safe zone exists, must be
considered strong candidates in the search for habitable planets.

\end{abstract}


\section{Introduction}

The discovery of hundreds of extrasolar planets has hurled the topic
of planetary studies into centerstage: planet formation, planetary
dynamics, planetary geology, etc.; among these studies, the question
of planet habitability is of great interest, even if no hard data
beyond Earth exists, yet. The quest for life on other planets started
long ago when in the 60's, Frank Drake used the 85-foot radius
telescope in West Virginia, hoping to detect an extraterrestrial
signal \citep{Drk61}. However, extraterrestrial life, if it indeed
exists in the neighborhood of the Sun, is most likely of a basic type
(e.g. microbial). Although a new emerging view is that planets very
different to Earth may have the right conditions for life, which
increases future chances of discovering an inhabited world
\citep{Seager13}, it is a good first step along this endeavor to find
out if conditions propitious to Earth-like planetary life (the only
type we know for sure so far) exists.

Even in the case of Earth, life exists in many diverse environments,
from scorching deserts to the eternal darkness of the ocean depths,
and even deep within Earth's crust. Confronted with this bewildering
diversity of environments, we must look for the most basic
ingredients, like liquid water. On the other hand, our observational
limitations to detect the most important habitability indicator: water
vapor on terrestrial-like exoplanets, reduces our possibilities of 
finding habitable planets. Up to now, we have been able to characterize 
habitable zones mainly around single stars, consequently, the quest for 
habitable worlds has been mostly limited to this type of stars. But
given the fact that most stars appear to be part of binary and
multiple systems, we must extend the habitable-zone concept to those
cases.

The condition for the existence of liquid water on the surface of a
planet \citep{bains04}, is the usual defining condition for what is
now called the circumstellar habitable zone (CHZ). \cite{Hart79}
studied the limits of the CHZ in late type stars and concluded that K
dwarfs have a narrow CHZ and later dwarfs have none at all. However,
other studies that include among other things, atmospheric radiative
transfer modeling, have come with a more optimistic and, rather
complicated outlook (e.g. \cite{Doyle98}). In this case, habitability
seems to be a very planet-specific matter \citep{Seager13}. For
example, the habitable zone calculations defined for a dry rocky
planet, with a minimum inner edge of about 0.5 AU, for a solar-like
host star (\cite{ZS13}), out to 10 AU for a planet with an H2
atmosphere and no interior energy, around a solar-like host star
\citep{PG11}, and even possibly out to free-floating habitable
planets, with no host star, for planets with thick H2 atmospheres
\citep{S99}.

For the case of planets in binary stellar systems, the usual radiative
condition used to define habitable zones around single stars must be
extended to the case of two sources of illumination whose relative
positions change in time. Additionally, a condition of dynamical
stability must be added, as the more complex and time varying
potential turns unstable whole swaths of phase space. For a planet to
be a possible adobe of life, both conditions must be satisfied, and
for a long time, enough for life to develop.

In the present work we will use \cite{Kopparapu2013} definition of
habitable zone around single stars and extend it to the case of binary
stars. As such, this is a simple definition of the habitable zone
based on stellar flux limits. Some of the more complicated (and
realistic) effects that pertain to particular situations, can be
incorporated into this simple criterion by the use of ``corrective
factors'' applied to the stellar fluxes that define the limits of the
habitable zones. An example of this is provided by e.g. \cite{KH13},
who introduce a factor called ``spectrally weighted flux''.

Studies like the previous ones, have been applied traditionally to
single stars or brown dwarfs, however, most low-mass main-sequence
stars are members of binary or multiple systems (\cite{DM91};
\cite{FM92}), and in particular in the Solar Neighborhood, the
fraction goes up to ~50\% (\cite{Abt83} \cite{Raghavan2010}). This
suggests that binary formation is at least as probable as single star
formation processes \citep{Math94}. Additionally, several types of
planets have been discovered in circumstellar orbits, even in close
binary systems, where the effect of the stellar companion might be of
great importance (\cite{DPL12}; \cite{CBL11}; \cite{MF10};
\cite{CU05}; \cite{ZM04}; \cite{HC03}; \cite{QM00}, and also in
circumbinary discs (\cite{DC11}; \cite{WO12}; \cite{OW12a};
\cite{OW12b}; \cite{SO13}). For a review on the subject see
\cite{H10}; and \cite{KH13}.

We provide in this work a general theoretical formulation to calculate
the equivalent of the \cite{Kopparapu2013} habitable zone for single
stars, but for binary systems (see section \ref{Sec:HabZone}). This
formulation is expressed in formulae and interpolating tables that the
reader can apply to any specific case. Our work also considers binary
systems with no restriction on the eccentricity of the binary orbits
and guarantees that orbits within the computed habitable zones remain
so, at all binary orbital phases. We only consider approximately
circular habitable zones. For the dynamical stability condition, we
use the work of \cite{PSA1} and \cite{PSA2}, who found the extent of
stable, non-intersecting orbits around binary systems, based on the
existence of sturdy structures in the extended phase-space of the
system (the so called “invariant loops”). We have then applied both
conditions to a sample of main sequence binary systems in the solar
neighborhood with known orbital parameters and present our
results. These constitute regions where we think it is most likely to
find planets suitable for life in these binary systems.

Several other interesting papers have been submitted on this subject
recently, for circular binary systems (\cite{Cuntz}), and for the
general case of elliptical binary systems (\cite{Eggl12};
\cite{Eggl13}; \cite{H10}; \cite{HS13}; \cite{KH13}; \cite{H14};
\cite{KaneHinkel}). However all these formulations take as the
stability criterion that of \citet{HW99}. The empirical approximation
of Holman \& Wiegert to the stability problem in binaries, was an
excellent and fundamental first approximation to the solution. Their
work was based on a detailed trial and error technique to find the
most stable (long-term) orbits in disks around binaries. The authors
defined as stable orbits mostly those that would keep in their orbits
for about $10^4$ binary periods (in general not enough to ensure life
emergency and development though). On the other hand, and as the
authors point out explicitly in their work, their formulation for
stability regions breaks toward higher eccentricity binaries due to
the rapidly increasing difficulty to recover stable orbits from this
methodology.

In this paper, we rely on the invariant loops theory, which by
definition are invariant structures in the extended phase-space of the
system. The existence of invariant loops guarantees the stability of
orbits for all times and not just during the integration span, as long
as the binary parameters (stellar masses, semi-major axis and orbital
eccentricity) do not change. Their stability is given by the constants
of motion that support them. The orbit integration is only used to
identify these structures in extended phase-space (for details see
\cite{PSA1}; \cite{MS00}; \cite{MS97}; \cite{A84};
\cite{LL92}). Because of its nature, this formulation has no
restrictions in eccentricity or mass ratio. With this criterion we
search for the intersection between the stable regions for planets and
a short, straightforward, and also general formulation for
habitability. Additionally, it is relevant to mention here, that the
use of the invariant loops criterion shows, for example in the
circumbinary disks case, important differences in the position of
stable regions \citep{PSA2} with respect to previous work, that should
be taken into account in habitability calculations. We explain this in
detail along this paper.

In the second part of this work we also provide a sample of binary
systems of the solar neighborhood (the whole sample of binary stars,
in the main sequence, with all orbital parameters known at the present
time in literature), with their habitable zones computed using the
approach in this work for habitability and planetary orbital stability.

This paper is organized as follows. In section \ref{Sec:HabZone} the
radiative condition for habitability is extended from the usual
circumstellar case, and a detailed description of the formulation used
to calculate the radiative safe zones is provided. In section
\ref{Sec:StabCrit} we present the dynamical stability condition for
habitability used in this work. \ref{Sec:partcases}, the combined
conditions are used to compute habitability zones for particular cases
and the construction of a table of binaries, with known orbital
parameters, for stars in the main sequence. Finally, our conclusions
are given in Section \ref{Sec:conclusions}.

\section{Radiative condition for habitable zones in binary stars}
\label{Sec:HabZone}

A common definition of the CHZ uses limits in the radiative stellar
flux at the planet (e.g. \cite{Kopparapu2013}):

\begin{equation}
I_{o}\le I(r)\le I_{i},
\label{eq:CHZdef}
\end{equation}
where the local flux is given by the stellar luminosity divided by
distance squared: $I = L/r^2$, and $I_{o}$, $I_{i}$, define the outer
and inner boundaries of the CHZ, respectively.

The CHZ thus defined is a thick spherical shell that surrounds the
star, whose thickness and size depend on the star's
luminosity. However, an important fraction of stars in the solar
neighborhood are part of binary systems. The fact that planets have
already been discovered within binary systems, makes it necessary to
extend the simple definition of the CHZ to the stellar binary case.

\begin{figure}
\begin{center}
\includegraphics[width=\linewidth]{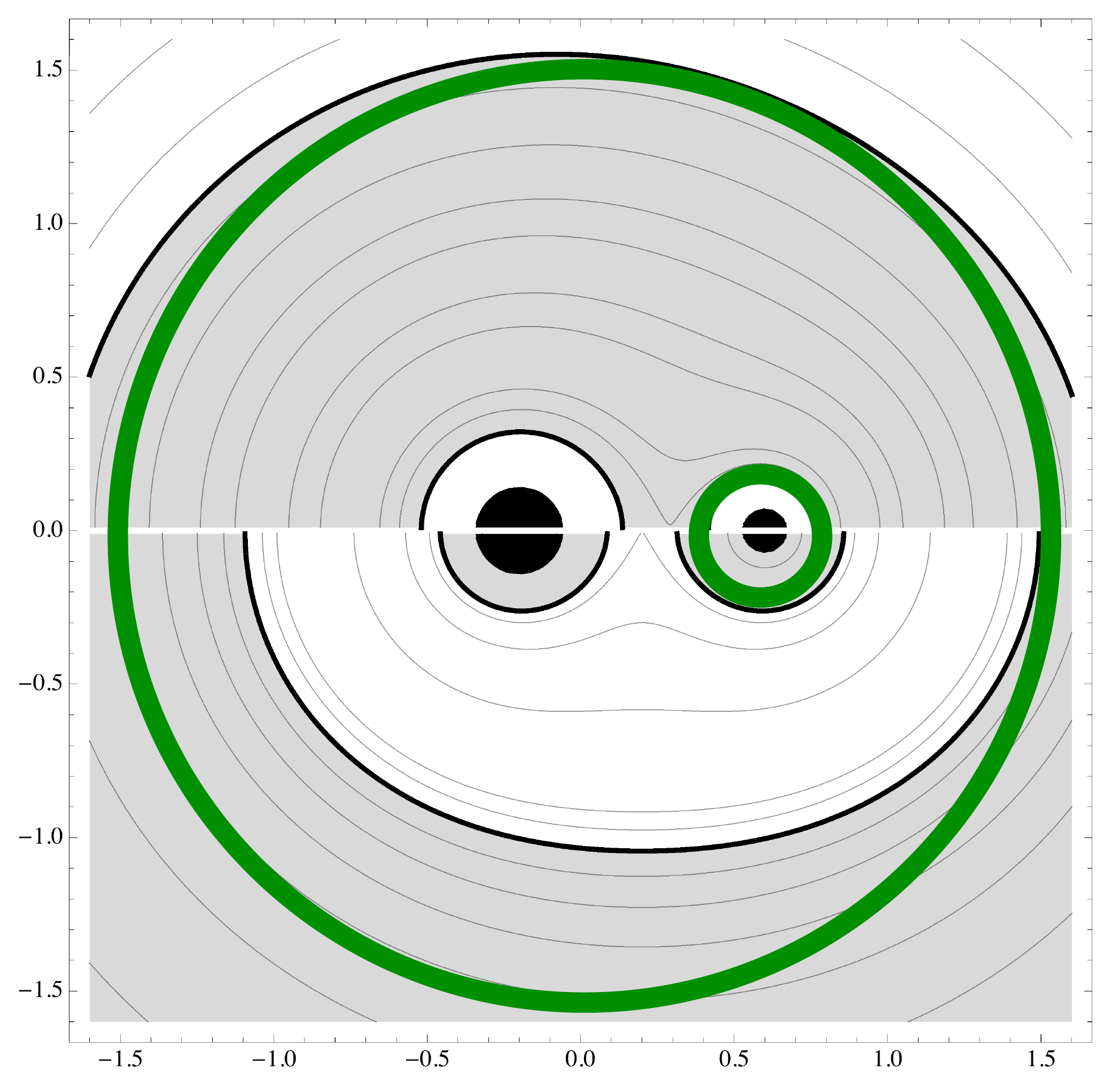}
\caption{Illustrative diagram for the concept of binary habitable zone. The stars are on the x-axis (black dots whose size is proportional to their luminosity      —upper half—, or mass —lower half—-). The upper part of the diagram shows the flux isophotes (equation 2) with the boundaries of the radiative safe zone shown with thick black lines and the zone itself shaded in gray. The lower part shows the equipotentials, with the circumbinary and circumstellar zones of dynamical stability sown in gray. Our definition of binary habitable zones (BHZ) is shown as the annular green region, where both conditions are met. In this case there is a circumbinary and a circumsecondary habitable zone.}
\label{fig:BHZ_diagram}
\end{center}
\end{figure}

In this section we extend the simple CHZ condition for single stars
given by the previous equation, to the case of stellar binary
systems. In this case we have two sources of luminosity at positions
that change with the binary orbital phase. It may be thought that in
this case it is simply a matter of applying the equation twice, once
for each star. However, this naive approach is not correct, since
there may be regions where, although within the individual CHZ for
each star, the combined irradiance of both stars may push the region
out of the combined habitable zone. Additionally, it is fundamental to add the condition of orbital stability, as both, the correct irradiance and orbital stable regions should have a non-empty intersection during the entire binary orbital phase, for planets to be able to exist within a proper binary habitable zone. In Figure \ref{fig:BHZ_diagram}, we present a schematic figure that shows the combined concept to construct habitable zones in a binary star: the radiative safe zone in gray circles (upper half of the diagram), and the stable regions for
orbits to settle down (lower half of the diagram). Notice how the demand that both conditions (radiative and dynamical) are met, severely restricts the resulting BHZ.

Since what matters is the total combined irradiance at a given point,
we should add the individual stellar fluxes in the CHZ condition given
by equation~(\ref{eq:CHZdef}), to arrive at the condition for the
radiative safe zone. The total stellar flux is given by:
\begin{equation}
I(x,y) = L_T\, \left[{
{{(1-\lambda_s)}\over{(x-r_p)^2 + y^2}} +
{{\lambda_s}\over{(x-r_s)^2 + y^2}}  }\right],
\label{eq:BinIrrad}
\end{equation}
here $L_T$ is the total binary luminosity and $\lambda_s$ is the
fractional contribution of the secondary star to it. $x$ and $y$ are
Cartesian coordinates in the binary orbital plane and $r_p$ and $r_s$
are the primary and secondary star distances to their barycenter. The
$x$-axis contains both stars.

To illustrate the concept of binary habitable zone, we show in fig. \ref{fig:BHZ_diagram} a schematic diagram that presents the radiative condition (upper half) and dynamical condition (lower half). The stars are the black dots on the x-axis, whose size is proportional to the assumed luminosity (upper half), or mass (lower half). In the upper half, isopleths of constant combined stellar flux are shown, with those corresponding to the boundaries of the raditaive safe zone shown with thick black lines. The raditaive safe zone is shaded in gray. Likewise, in the lower half of the diagram, equipotentials are shown with the circumbinary and two circumstellar dynamical safe zones in gray. Notice that the saddle point between the stars for isopleths and equipotentials does not coincide, as the former is set by the relative luminosities, whereas the latter by their mass ratio. 

We define our binary habitable zones (BHZ) as the annular regions (shown in green) where both conditions are met. For the circumbinary habitable zone, it is an annular region centered at the barycenter of the system. For the     circumstellar habitable zones, they are annular regions centered in the   corresponding star. Notice that in this
particular example there is a circumbinary and a circumsecondary
habitable zones, but not a circumprimary. In this section we will tackle the radiative condition alone, leaving the condition of dynamical stability for the next section and the combined effect for section 4.

\begin{figure}
\begin{center}
\includegraphics[width=\linewidth]{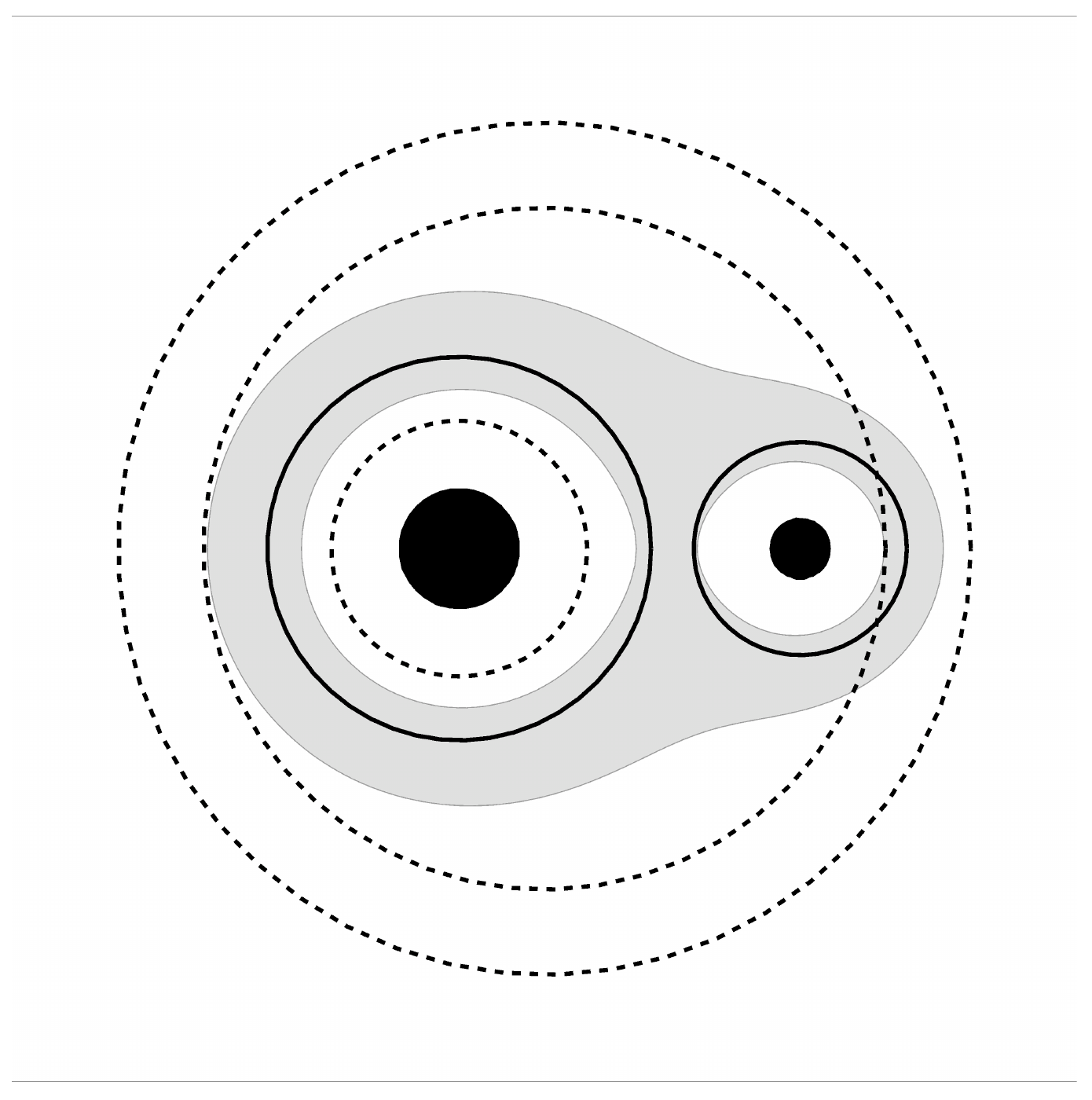}
\caption{Safe (continuous) and unsafe (dashed) planetary orbits. The radiative safe zone is the gray region. The planetary orbits are simply circles centered on either star, or the center of mass of the system. In this case no circumbinary safe orbits are possible.}
\label{fig:BHZillust}
\end{center}
\end{figure}

The edges of the radiative safe zone are set by two stellar flux
isopleths and the resulting shape is more complicated that that of the
CHZ. Figure (\ref{fig:BHZillust}) illustrates a particular example:
the gray region is the radiative safe zone. Dashed lines indicate
possible planetary orbits that are not fit for life, while solid lines
indicate safe orbits (we have approximated the planetary orbits as
circles centered in either star, or the barycenter). Notice that for
an orbit to be safe, it must remain within the radiative safe zone at
all times. In this particular case, there are safe circumprimary and
circumsecondary planetary orbits, but no circumbinary ones.

In addition to the above complication, the separation between the
stars will vary for the general case of binaries with elliptical
orbits. First we will tackle the simpler case of circular orbit
binaries.

\subsection{Binaries in circular orbits}

The star distances to the barycenter are given by::
\[
r_p = -(m_s/M) r_{12}, \quad r_s = +(m_p/M) r_{12},
\]
where the $p$ and $s$ subindexes refer to the primary and secondary
stars, $m$ are the individual stellar masses and $M$ is the total
mass. Finally, $r_{12}$ is the constant distance between the stars.

If we take the interstellar distance and total mass as units of length
and mass, we can write the previous relation as:
\begin{equation}
\eta_p = -q, \quad \eta_s = +(1-q),
\label{eq:qshift}
\end{equation}
where $\eta$ is dimensionless distance and $q=m_s/M$ is the mass
fraction due to the secondary.

A similar scaling in luminosity can be accomplished adopting $L_T$ as
its unit. The dimensionless combined stellar flux is then:
\begin{equation}
{\cal I}(\eta_x,\eta_y) = 
{{(1-\lambda_s)}\over{(\eta_x+q)^2 + \eta_y^2}} +
{{\lambda_s}\over{(\eta_x+q-1)^2 + \eta_y^2}} ,
\label{eq:ScaledBinIrrad}
\end{equation}
where $(\eta_x, \eta_y)$ are the corresponding dimensionless Cartesian
coordinates. Notice that ${\cal I}$ is completely set by the
luminosity and mass ratios, its unit is $L_T/r_{12}^2$.

\subsubsection{The critical flux isopleth}

\begin{figure}
\begin{center}
\includegraphics[width=\linewidth]{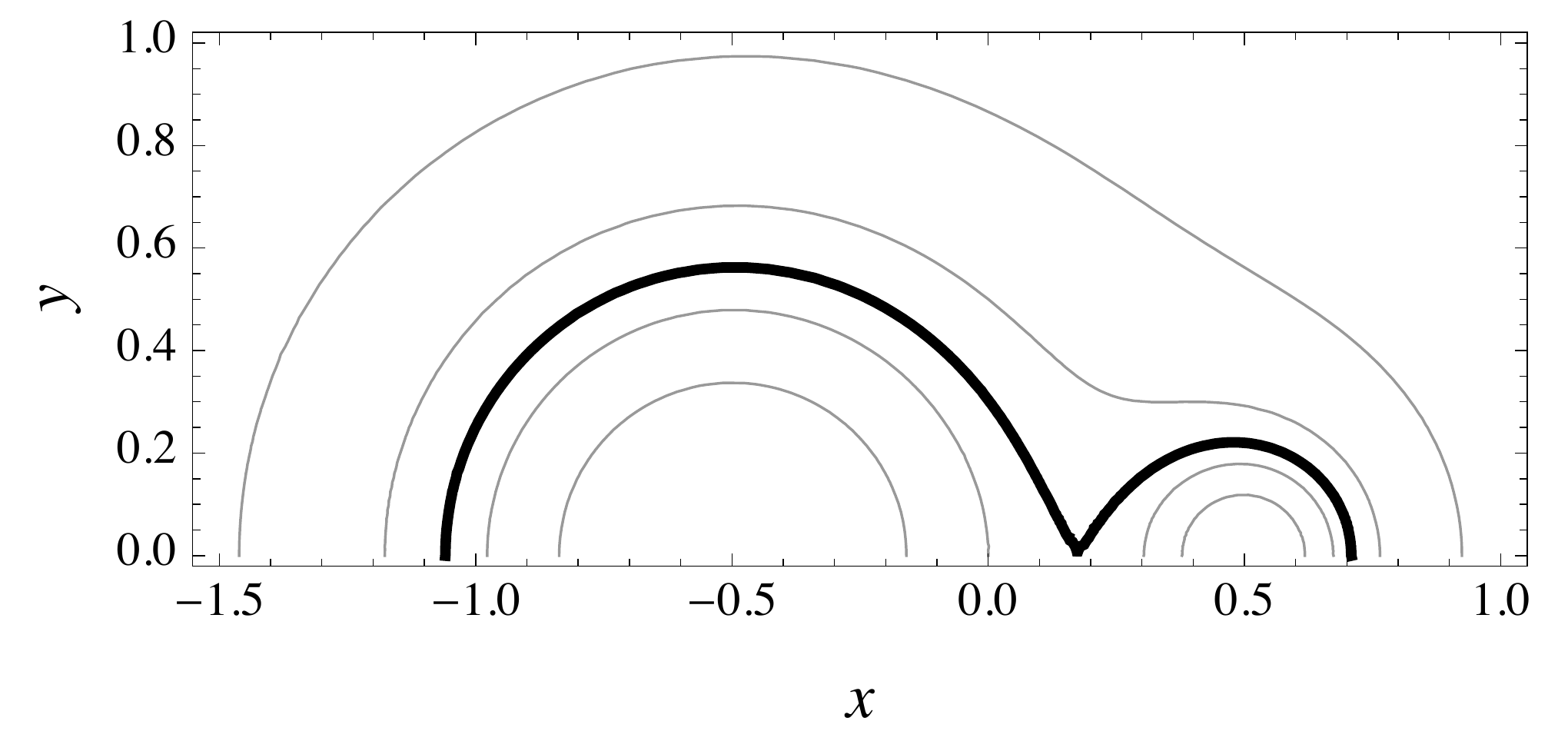}
\caption{Various flux isopleths are shown for a binary with $q=0.5$,
  $\lambda=0.1$. The thick line is the critical isopleth. The plot
  origin is at the barycenter frame and only the positive $y$ axis is
  shown.}
\label{fig:Isopleths}
\end{center}
\end{figure}

The shape of the ${\cal I}$--isopleths changes from a single
(circumbinary) to a double (circumstellar) contour at the {\it
  critical isopleth} (see figure~\ref{fig:Isopleths}). Its value is
given by:

\begin{equation}
\Ic(\lambda) = {{(1-\lambda_s)^{1/3} + \lambda_s^{1/3}}\over{\left[{(1-\lambda_s)^{2/3}+\lambda_s^{2/3}-\lambda_s^{1/3}(1-\lambda_s)^{1/3}}\right]^2}}
\label{eq:Ic}
\end{equation}

\begin{figure}
\begin{center}
\includegraphics[width=\linewidth]{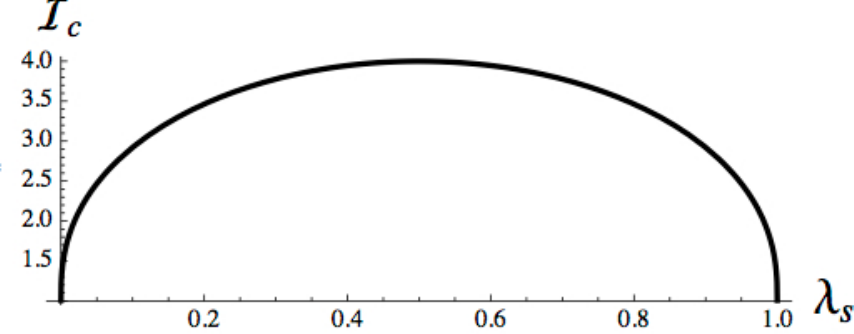}
\caption{Critical isopleth flux value as a function of the secondary
  star luminosity fraction. Notice that $\Ic$ is symmetric with
  respect to $\lambda_s=0.5$.}
\label{fig:Ic}
\end{center}
\end{figure}

Figure~(\ref{fig:Ic}) shows the value of the critical flux isopleth as
a function of the secondary luminosity fraction. Notice that ${\cal I}_c$ 
does not depend on the mass ratio $q$. This is because the latter only
shifts the isopleths on the orbital plane (see
equations~\ref{eq:qshift}). The curve is also symmetric with respect
to $\lambda_s=1/2$. This is because $\lambda$ can refer to either star
(the more massive star is not necessarily the most luminous one). So
we drop the $s$ subindex from now on.

\subsubsection{The three different radiative zone configurations}
\label{HZ}

\begin{figure}
\begin{center}
\includegraphics[width=\linewidth]{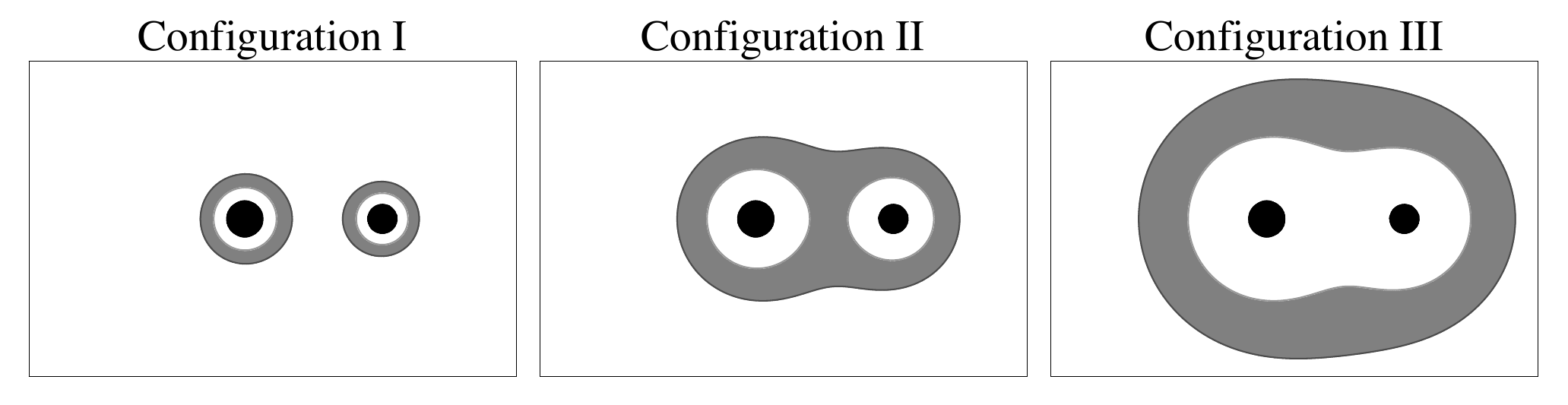}
\caption{The three different radiative zone configurations. The illustration is
  for a binary with $\lambda=0.4$.}
\label{fig:BHZcases}
\end{center}
\end{figure}

Depending on the value of the dimensionless stellar fluxes that define
the CHZ (equation~\ref{eq:CHZdef}) with respect to the critical
isopleth flux, the radiative safe zone may have three different
configurations:

\begin{description}
\item[{Configuration I:} ($\Ic < \Io$).] This corresponds to 2
  separate circumstellar radiative safe zones.
\item[{Configuration II:} ($\Io < \Ic < \Ii$).] This corresponds to a
  combined case, where we find a circumbinary radiative safe zone with
  two inner holes around the stars.
\item[{Configuration III:} ($\Ii<\Ic$).] This corresponds to a single
  radiative safe zone with a single inner hole.
\end{description}
These configurations are illustrated in figure~\ref{fig:BHZcases}.

In any of these configurations, safer planetary orbits are the ones
better contained within the safe radiative zone. In the most strict
sense, this means that we must consider their shape, which introduces
an added complexity we will not get into. Since stable orbital
planetary configurations in binaries have in general low
eccentricities (smaller to $\sim$0.3, \cite{PSA1}), we will define a 
radiative Binary Habitable Zone (rBHZ) as the largest circular annular region
that can fit entirely within the radiative safe zone. We consider two
cases: circumstellar and circumbinary zones. Orbits that are safe for
life will lay most likely within these zones. We should remember that
these zones are defined by the radiative condition only. In next
section we will consider the dynamical condition and in section
\ref{Sec:partcases} we will combined both conditions to define the
combined binary habitable zone.

Now, a problem arises because the flux isopleths are not circles (see
figure~\ref{fig:Isopleths}), particularly close to the critical
isopleth. Our definition of BHZ implies that the largest safe orbit
must be entirely inscribed within the flux isopleth that defines the
outer boundary of the safe zone. Similarly, the smallest safe orbit
must be entirely circumscribed outside the inner boundary flux
isopleth. Note that this condition is not over the planetary orbit
itself but just in order to fix the boundaries for habitability, in
such way that it is possible to ensure that a planet will remains all the
time inside this zone.

The study can be separated depending on the configuration of the radiative 
safe zones, we now examine these conditions for each case.

\subsubsection{Configuration I ($\Ic < \Io$)}
In this case we have two separate circumstellar habitable zones. The
orbital radius of the largest safe planetary orbit is given by the
smallest distance between the outer boundary isopleth and the
corresponding star. This ensures that the orbit will remain within the
outer boundary isopleth.

Similarly, for the radius of the smallest, circumstellar safe orbit,
we must now consider the maximum distance from either star to the
corresponding inner boundary flux isopleth, so the orbit remains
outside it at all times.

\begin{figure}
\begin{center}
\includegraphics[width=\linewidth]{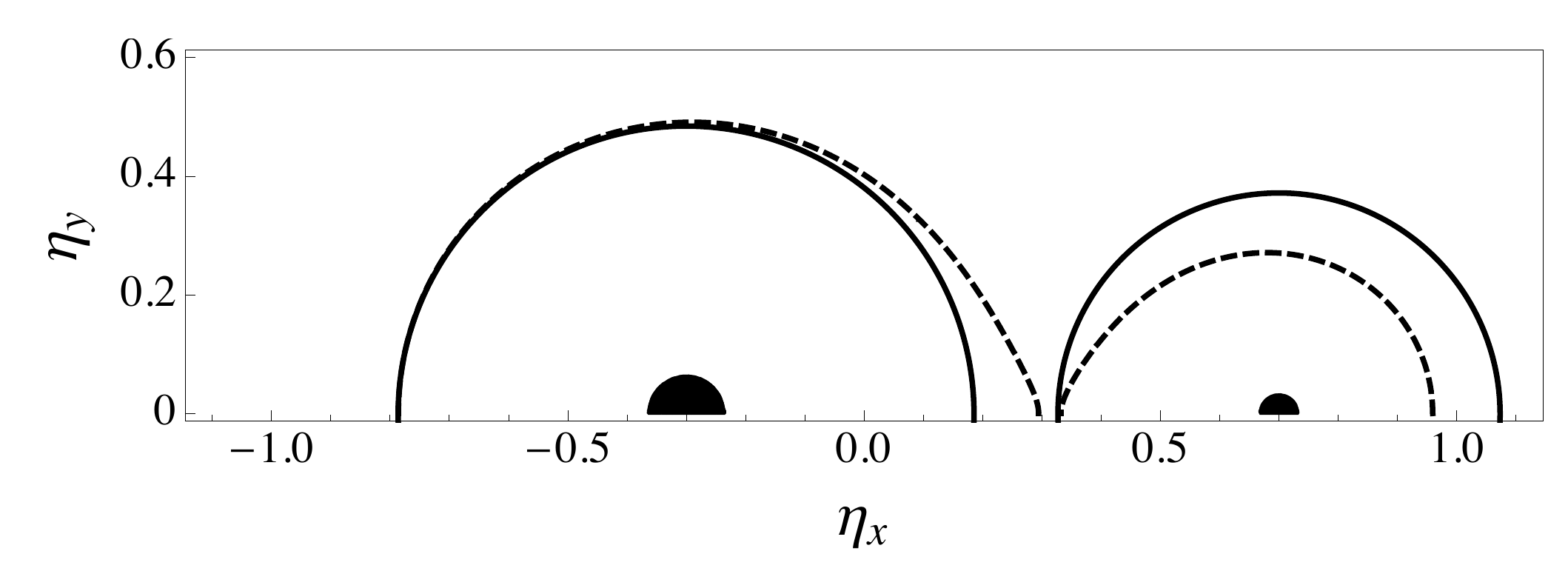}
\caption{Circumstellar boundaries of the rBHZ (configuration I). The
  boundary isopleth is the dashed line. If this isopleth corresponds
  to the outer boundary, the largest safe orbit touches the isopleth
  on its $\eta_x$ intersection opposite the other star (left solid
  circle). If the isopleth is the inner boundary, the smallest safe
  orbit touches the isopleth at the opposite $\eta_x$ intersection
  (right solid circle). The case shown is $q=0.3$, $\lambda=0.2$ and
  the isopleth depicted is $1.0045\Ic$. The primary star, which in
  this example is also the most luminous, is the one on the left.}
\label{fig:cstellar}
\end{center}
\end{figure}

The circumstellar isopleths are elongated along the $\eta_x$ axis in
the direction of the other star, while they are squashed in the
opposite direction. This means that both, the outer and inner
circumstellar orbital radii ($r_{os}$ and $r_{is}$, respectively)
correspond to the respective boundary isopleth intersections with the
$\eta_x$ axis (see figure \ref{fig:cstellar}).

\begin{figure}
\begin{center}
\includegraphics[width=\linewidth]{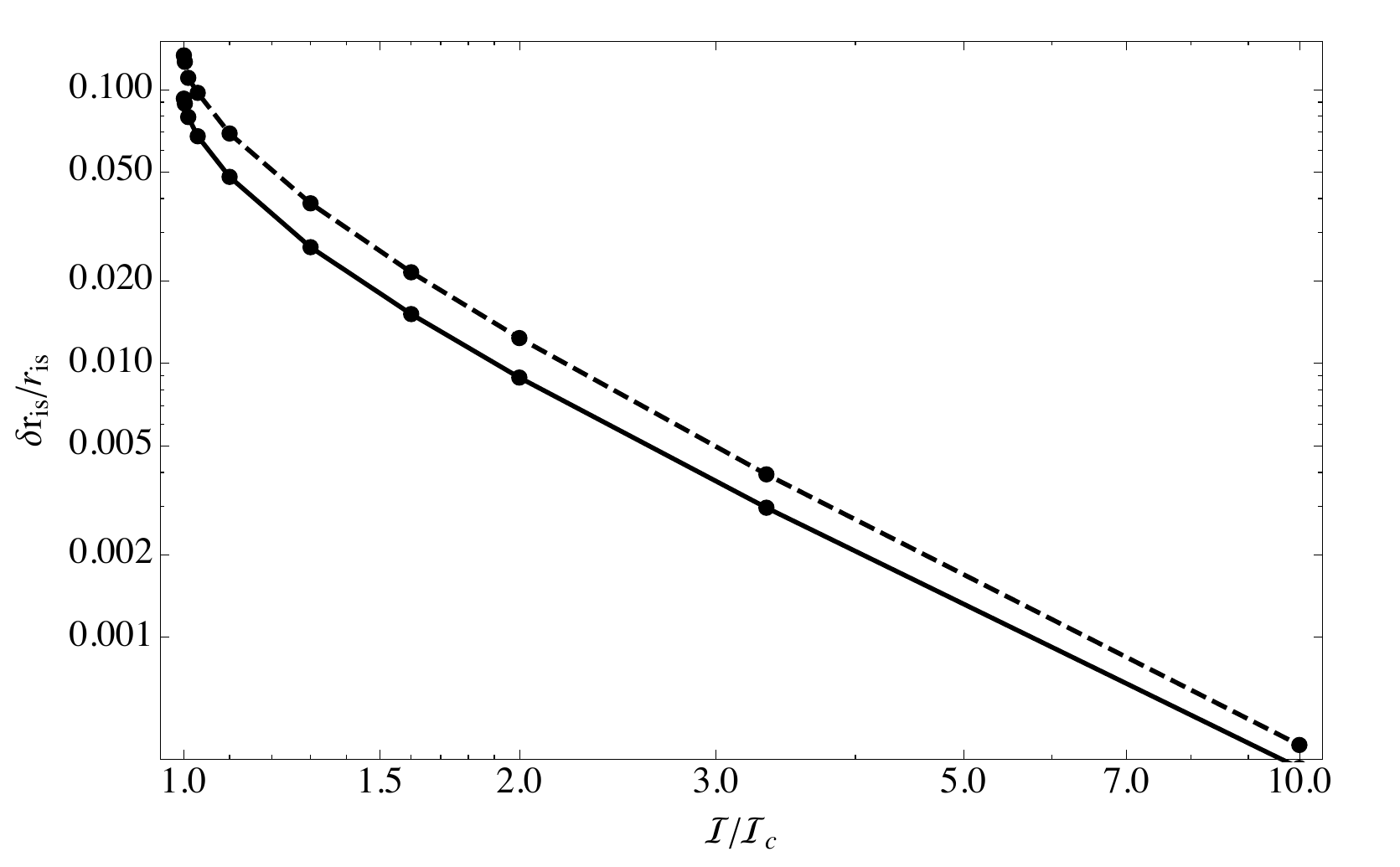}
\caption{Fractional error made using the CHZ criterion for a single
  star, instead of $r_{is}$, as a function of the inner boundary flux
  value (in units of the critical flux value). The solid line is for
  the most luminous star ($\lambda=0.9$), while the dashed line is for
  the less luminous star ($\lambda=0.1$).}
\label{fig:risApprox}
\end{center}
\end{figure}

From equation~(\ref{eq:ScaledBinIrrad}), setting $\eta_y=0$ and
putting the star in question at the coordinate origin, we arrive at
the following polynomial:

\begin{equation}
\Ib r_s^4 - 2\Ib r_s^3 + (\Ib-1) r_s^2 + 2\lambda r_s - \lambda = 0,
\label{eq:rs}
\end{equation}
where $\Ib$ is the dimensionless flux value that defines the rBHZ
boundary and $r_s$ is either, the outer circumstellar orbital radius
$r_{os}$, or the inner one $r_{is}$, depending on the solution we
choose. Unfortunately, although 4--degree polynomials can always be
solved by radicals, the solution in this case is quite cumbersome. We
have decided instead, to list in table~\ref{tab:rs} some
solutions. The first value at each entry corresponds to $r_{os}$
(smallest value), the second to $r_{is}$. We list results from the
critical isopleth up to thrice its value only, because for larger
values, the circumstellar boundaries are very close to the star and
the single star CHZ criterion (equation~(\ref{eq:CHZdef})) is
sufficient. This is shown in figure~\ref{fig:risApprox}, where the
fractional error made in using the CHZ criterion individually for each
star, instead of the true value obtained by solving
equation~(\ref{eq:rs}), is shown as a function of the flux value for
the inner boundary. We can see that for values larger than those
listed in table~\ref{tab:rs}, the error drops well below 1\%.

\begin{table*}
\begin{minipage}{126mm}
  \caption{Some values of the radius of the largest/smallest safe
    circumstellar orbits $r_{s}$ (real solutions to
    equation~(\ref{eq:rs}) for $\Ib>\Ic$)}
\label{tab:rs}

\begin{tabular}{lccccccccc}
 \hline
 $\Ib/\Ic$ & $\lambda=0.1$ & $\lambda=0.2$ & $\lambda=0.3$ & $\lambda=0.4$ & $\lambda=0.5$ & $\lambda=0.6$ & $\lambda=0.7$ & $\lambda=0.8$ & $\lambda=0.9$ \\
  \hline
  1           & 0.208/0.343 & 0.260/0.387 & 0.299/0.430 & 0.333/0.466 & 0.366/0.500 & 0.400/0.534 & 0.439/0.570 & 0.487/0.614 & 0.559/0.675 \\
  1.001   & 0.208/0.316 & 0.260/0.378 & 0.299/0.421 & 0.333/0.457 & 0.366/0.491 & 0.400/0.525 & 0.439/0.561 & 0.487/0.605 & 0.559/0.667 \\
  1.002   & 0.208/0.313 & 0.260/0.374 & 0.298/0.417 & 0.333/0.453 & 0.366/0.487 & 0.400/0.521 & 0.439/0.557 & 0.486/0.601 & 0.558/0.663 \\
  1.005   & 0.208/0.306 & 0.259/0.367 & 0.298/0.410 & 0.332/0.446 & 0.365/0.480 & 0.399/0.513 & 0.438/0.550 & 0.486/0.593 & 0.558/0.656 \\
  1.01     & 0.207/0.298 & 0.258/0.359 & 0.297/0.402 & 0.331/0.438 & 0.364/0.471 & 0.398/0.505 & 0.437/0.541 & 0.484/0.585 & 0.556/0.648 \\
  1.02     & 0.206/0.288 & 0.257/0.348 & 0.296/0.390 & 0.329/0.426 & 0.362/0.459 & 0.396/0.493 & 0.435/0.529 & 0.482/0.573 & 0.553/0.636 \\
  1.05     & 0.202/0.269 & 0.253/0.327 & 0.291/0.368 & 0.324/0.403 & 0.357/0.436 & 0.390/0.469 & 0.428/0.506 & 0.475/0.549 & 0.545/0.612 \\
  1.1       & 0.197/0.248 & 0.246/0.305 & 0.284/0.345 & 0.316/0.379 & 0.348/0.411 & 0.381/0.444 & 0.418/0.479 & 0.464/0.522 & 0.533/0.584 \\
  1.2       & 0.187/0.223 & 0.234/0.277 & 0.271/0.314 & 0.302/0.347 & 0.333/0.377 & 0.364/0.409 & 0.400/0.443 & 0.444/0.485 & 0.510/0.546 \\
  1.5       & 0.164/0.181 & 0.207/0.228 & 0.240/0.262 & 0.269/0.291 & 0.296/0.319 & 0.325/0.347 & 0.357/0.378 & 0.396/0.417 & 0.456/0.473 \\
  2          & 0.139/0.147 & 0.177/0.187 & 0.206/0.216 & 0.231/0.242 & 0.255/0.266 & 0.280/0.291 & 0.308/0.318 & 0.343/0.352 & 0.394/0.402 \\
  3          & 0.112/0.115 & 0.143/0.147 & 0.167/0.171 & 0.187/0.191 & 0.207/0.211 & 0.228/0.232 & 0.251/0.255 & 0.279/0.283 & 0.321/0.324 \\
  \hline
\end{tabular}
\end{minipage}

\end{table*}

\subsubsection{Configuration III ($\Ii<\Ic$)}

The considerations that define the largest and smallest boundaries are
the same as for the previous case, except that this time we are
dealing with boundary isopleths that surround both stars and the
relevant extremal distances are with respect to the binary barycenter.

In this case there are two additional difficulties: the position of
the isopleths with respect to the barycenter depends on the mass ratio
too and their form, close to the critical one, is nowhere near
circular.

\begin{figure}
\begin{center}
\includegraphics[width=\linewidth]{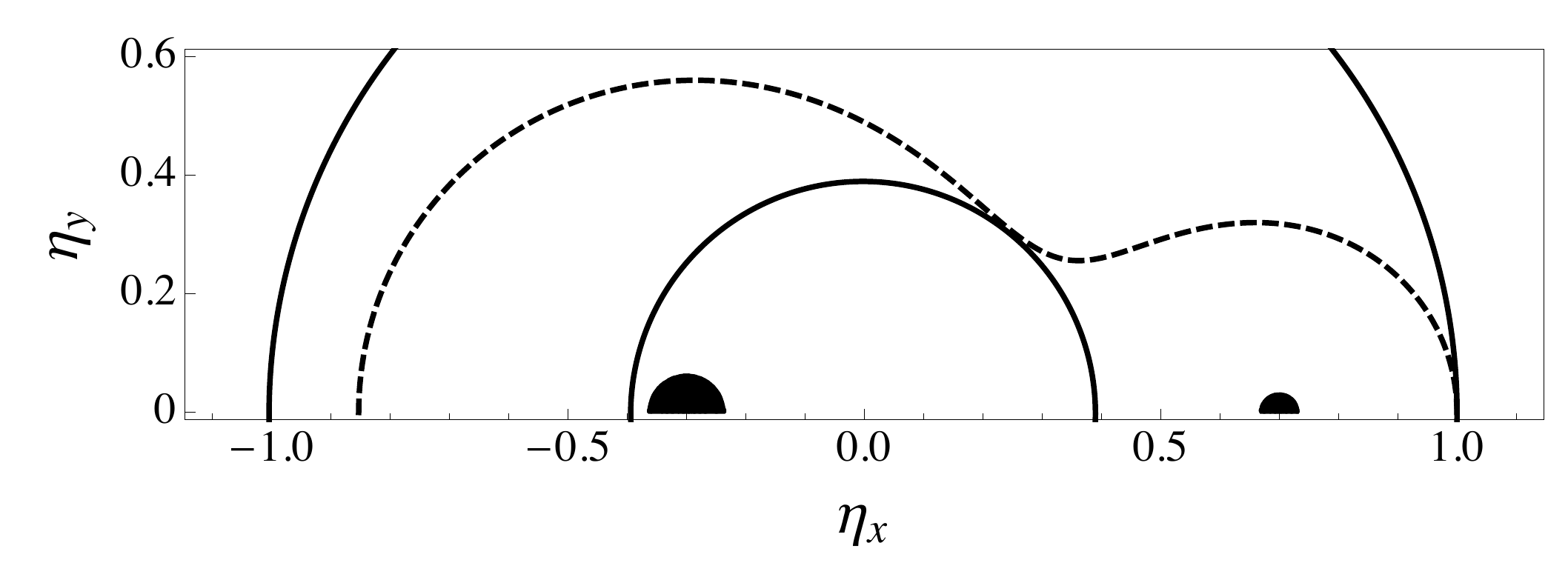}
\caption{Circumbinary boundaries of the rBHZ (configuration III). The
  boundary isopleth is the dashed line and the barycenter is at the
  coordinate origin. If this isopleth corresponds to the outer
  boundary, that largest safe orbit touches the isopleth at the point
  where the barycenter to isopleth distance is minimum (inner
  circle). If the isopleth is the inner boundary, the smallest safe
  orbit touches the isopleth at the $\eta_x$ intersection farthest
  away from the barycenter (outer circle). The binary parameters are
  the same as those in figure (5), except that the isopleth depicted
  is $0.779\Ic$}
\label{fig:cbinary}
\end{center}
\end{figure}

For the outer boundary, the relevant distance is that from the
barycenter to the closest point along the boundary isopleth. In
general, this point is not along either axis (see
figure~\ref{fig:cbinary}).

From equation~(\ref{eq:ScaledBinIrrad}), we obtain the
barycenter--isopleth distance. Minimizing this function, we get the
orbital radius of the largest safe orbit:

\begin{equation}
r_{ob}(q,\lambda,\Io)={1\over{\Io}} \sqrt{{A_1 + A_2}\over{2q f_1(q)}},
\label{eq:rob}
\end{equation}
where $r_{ob}$ denotes the outer circumbinary orbital radius and $\Io$
is the flux value that defines the outer boundary isopleth (in this
case ${\cal I}_o<{\cal I}_c$). The auxiliary quantities are given by:

\begin{equation}
A_1 = {2[\Io q\faq]^2} + {\Io f_2^2(q)\sqrt{\lambda q\fal\faq}} ,
\end{equation}
\begin{equation}
A_2 = \Io q \faq \Bigl\lbrack \sqrt{{\lambda\fal}\over{q\faq}} - 2[q\fbl+\fal] \Bigr\rbrack,
\end{equation}
and 
\begin{equation}
f_1(x) = x-1, \quad f_2(x) = 1-2x
\end{equation}

Equation~(\ref{eq:rob}) is symmetric around $\lambda=1/2$. This is
expected, since a reflexion around the $\eta_y$ axis of
figure~(\ref{fig:cbinary}) leaves the shortest barycenter to isopleth
distance unchanged.

Notice that the radius of the largest safe circumstellar orbit does
not depend on $q$, whereas that of the circumbinary orbit does. As we
mentioned, this is because the position of the isopleths is fixed with
respect to the stars, but not with respect to the barycenter
(equation~\ref{eq:qshift}).

For the inner boundary, the proper orbital radius is the distance from
the barycenter to the farthest away $\eta_x$ axis intersection of the
boundary isopleth (figure~\ref{fig:cbinary}). The easiest way to find
this distance is to solve equation~(\ref{eq:ScaledBinIrrad}) for
$\eta_y=0$, which leads to the same polynomial as in case~I
(equation~\ref{eq:rs}). But now ($\Ib<\Ic$), the polynomial has only
two real roots, which are the distances from either star to the
boundary isopleth ($r_\star$). To find the final inner circumbinary
orbital radius, we now add the respective star to barycenter distance
(equation~\ref{eq:qshift}) to each root and compare the resulting
barycenter to isopleth distances. The largest one is $r_{ib}$.

\begin{figure}
\begin{center}
\includegraphics[width=\linewidth]{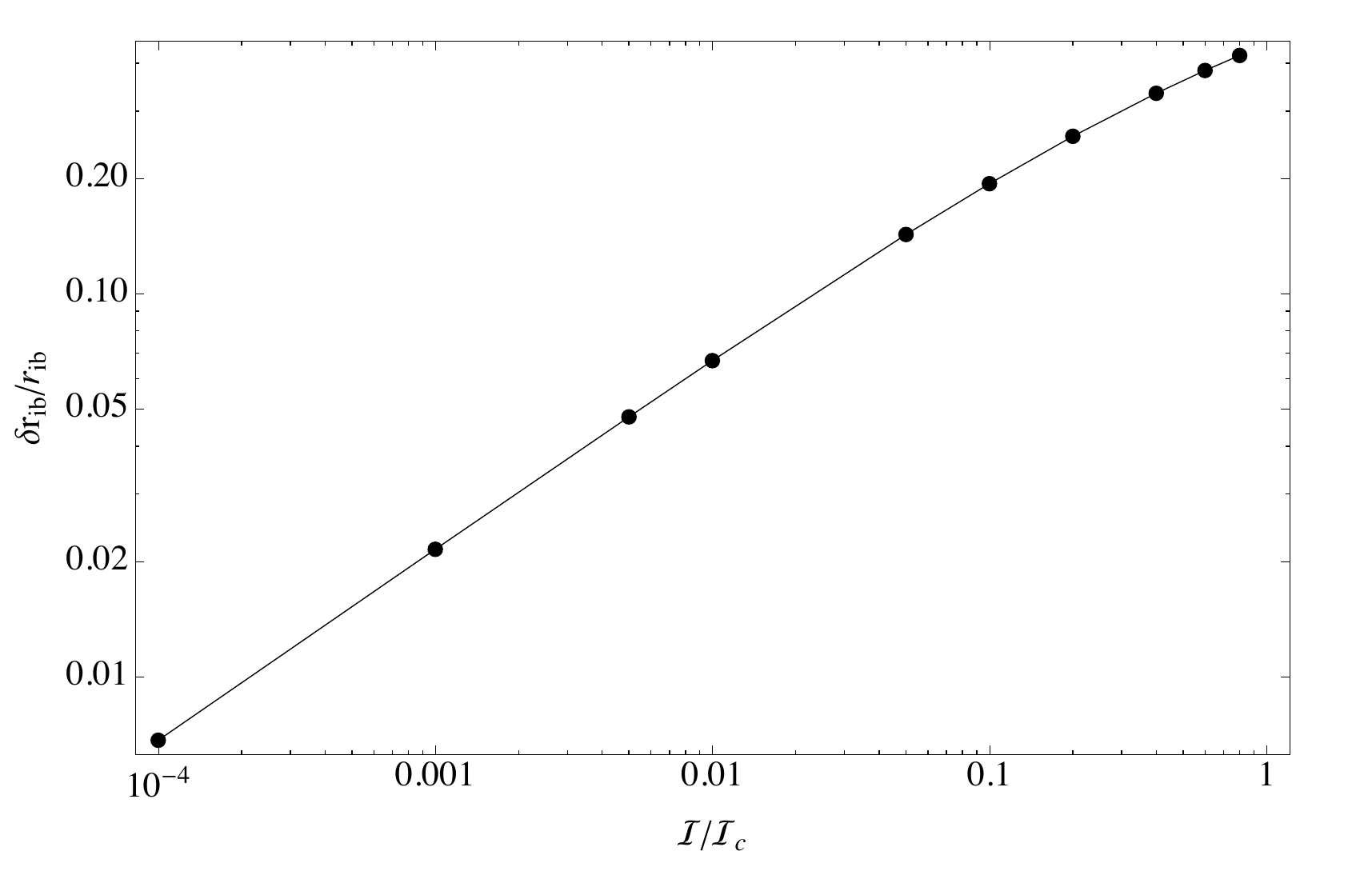}
\caption{Fractional error made using the CHZ criterion for a single
  star at the barycenter with the total binary luminosity, instead of
  $r_{ib}$, as a function of the inner boundary flux value (in units
  of the critical flux value). The $\lambda=0.9$ case is shown.}
\label{fig:ribApprox}
\end{center}
\end{figure}

As before, the solution is rather complicated and some values of
$r_\star$ are shown in table~\ref{tab:rib}. For distant boundaries
(small values of $\Ii/ \Ic$), the isopleths become circular and
converge to the CHZ criterion applied using the total binary
luminosity. This can be seen in figure~(\ref{fig:ribApprox}), where
the fractional error made using the CHZ instead of the rBHZ criterion
is plotted for the $\lambda=0.9$ case. Again, for values beyond those
listed in table~\ref{tab:rib} the error is very small.

\begin{table*}
\begin{minipage}{126mm}
  \caption{Some values of the star to outer boundary isopleth
    $r_{\star}$ (negative real solution to equation~(\ref{eq:rs}) for
    $\Ib<\Ic$)}
\label{tab:rib}

\begin{tabular}{lccccccccc}
 \hline
 $\Io/\Ic$ &$\lambda=0.1$ & $\lambda=0.2$ &  $\lambda=0.3$ &  $\lambda=0.4$ & $\lambda=0.5$ & $\lambda=0.6$ &  $\lambda=0.7$ &  $\lambda=0.8$ &  $\lambda=0.9$ \\
  \hline
 0.99     & 0.210 & 0.261 & 0.300 & 0.335 & 0.368 & 0.403 & 0.441 & 0.489 & 0.562  \\
 0.90     & 0.222 & 0.276 & 0.316 & 0.352 & 0.387 & 0.423 & 0.463 & 0.514 & 0.589  \\
 0.80     & 0.239 & 0.295 & 0.338 & 0.375 & 0.412 & 0.450 & 0.492 & 0.546 & 0.626  \\
 0.60     & 0.287 & 0.349 & 0.397 & 0.439 & 0.480 & 0.523 & 0.571 & 0.632 & 0.723  \\
 0.40     & 0.379 & 0.446 & 0.500 & 0.549 & 0.597 & 0.648 & 0.705 & 0.778 & 0.888  \\
 0.20     & 0.635 & 0.695 & 0.754 & 0.812 & 0.872 & 0.938 & 1.014 & 1.111 & 1.262  \\
 0.10     & 1.084 & 1.101 & 1.151 & 1.213 & 1.282 & 1.363 & 1.461 & 1.590 & 1.795  \\
 0.05     & 1.796 & 1.736 & 1.759 & 1.814 & 1.889 & 1.984 & 2.106 & 2.275 & 2.552  \\
 0.01     & 4.978 & 4.624 & 4.512 & 4.508 & 4.574 & 4.701 & 4.900 & 5.211 & 5.769  \\
 0.005   & 7.392 & 6.833 & 6.623 & 6.571 & 6.624 & 6.767 & 7.017 & 7.427 & 8.188  \\
 0.001   & 17.61 & 16.20 & 15.59 & 15.34 & 15.34 & 15.54 & 15.99 & 16.80 & 18.40  \\
 0.0005 & 25.27 & 23.24 & 22.33 & 21.93 & 21.88 & 22.13 & 22.72 & 23.84 & 26.07  \\
 0.0001 & 57.60 & 52.93 & 50.76 & 49.75 & 49.51 & 49.95 & 51.16 & 53.53 & 58.40   \\
 \hline
\end{tabular}
\end{minipage}

\end{table*}

In this case it may be that $r_{ib}>r_{ob}$, which means that no safe
circular orbits centered in the barycenter exist. However, a
non--circular and sufficiently elongated orbit may remain within the
rBHZ, but this is beyond this study.

\subsubsection{Configuration II ($\Io < \Ic < \Ii$)}
This is the most complicated configuration to compute, since we now
have the possibility of both, circumstellar and circumbinary safe
zones. This is a mixture of the previous two cases and they have to be
dealt separately.

For the circumstellar zone, we use the procedure used for case~I, with
$\Ib=\Ii$, to interpolate values of $r_{is}$ for each star from
table~\ref{tab:rs}. For the outer edge, we use the procedure of
case~III with $\Ib=\Io$, to obtain from equation~(\ref{eq:rob}) a
value that we will interpret as $r_{os}$. Notice that for the outer
circumstellar edge we are using an isopleth that is not
circumstellar. This is because a safe circumstellar orbit may reach
beyond the critical isopleth\footnote{We should remember that the
  Roche lobes, which are the regions of dynamical influence of each
  star, are not the same as the circumstellar regions defined by the
  critical flux isopleth.} (see figure~\ref{fig:example2}). It is
obvious that if the resulting $r_{os}$ reaches all the way to the
inner edge of the circumstellar safe zone of the other star, then the
smaller of the two distances should be taken, since a safe
circumstellar orbit should not get into the unsafe region of the other
star.

\begin{figure}
\begin{center}
\includegraphics[width=\linewidth]{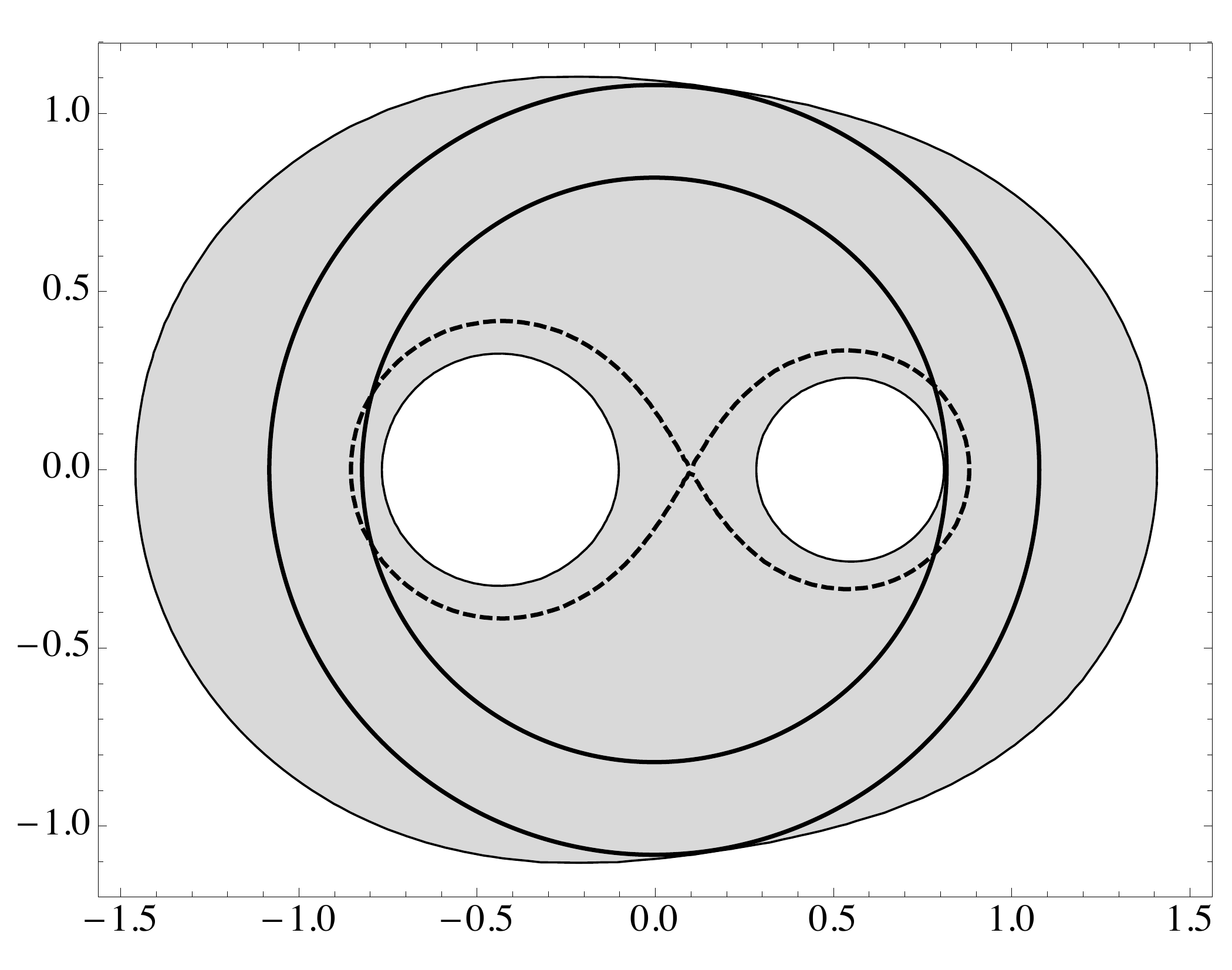}
\caption{Circumbinary safe orbits in configuration~II. The dashed line
  is the critical isopleth and the shaded area is the radiative safe
  zone. The extremal safe circumbinary orbits, which define the
  boundaries of the circumbinary rBHZ are shown (continuous line
  circles). The example shown is for $q=0.44$, $\lambda=0.375$ and
  $\Io/ \Ic = 0.179$, $\Ii/ \Ic=1.592$.}
\label{fig:example0}
\end{center}
\end{figure}

For the circumbinary region, we apply the procedure of case~III with
$\Ib=\Io$ for the outer edge. For the inner edge we have to deal again
with the possibility of safe orbits that cross the critical isopleth
(see figure~(\ref{fig:example0})). The inner circumbinary orbital
radius is given by the largest of the two barycenter to inner boundary
isopleth $\eta_x$ intersections farther away in the direction to each
star. These distances are given by the individual $r_{is}$ for each
star (which we computed already), plus their respective star distance
to the barycenter.

\subsubsection{Procedure to obtain the habitability region}

Tables~\ref{tab:rs} and \ref{tab:rib}, together with
equation~(\ref{eq:rob}) and auxiliary relations, define the orbital
radii of the largest and smallest circumstellar or circumbinary safe
edges.

The detailed procedure to find the limits of the rBHZ is then as follows:

\begin{enumerate}

  \item From the binary components individual masses and luminosities,
    obtain the dimensionless mass fraction $q$ and luminosity fraction
    $\lambda$, the latter for both stars.

  \item Get specific boundary radiative stellar flux values $I_o$,
    $I_i$ (e.g. \cite{Kopparapu2013}) and convert them to the
    dimensionless system we use here: $\Ii$, $\Io$ (normalized to
    total binary luminosity per square semimajor axis).

  \item Given $\lambda$ (either star), use equation~(\ref{eq:Ic}) to
    find the value of the critical flux for this case: $\Ic$.
  
  \item If $\Io>\Ic$, proceed to configuration~I; if $\Ii<\Ic$,
    proceed to configuration~III; else, go to configuration~II.

  \item {\bf Configuration I}: Two separate radiative circumstellar habitable
    zones. Use table~\ref{tab:rs} to interpolate the $r_{os}$ and
    $r_{is}$ edges radii for each star using the appropriate value of
    $\Ib$.

  \item {\bf Configuration II}: One radiative circumbinary habitable zone with
    two circumstellar inner edges. We have to solve the circumstellar
    and circumbinary zones separately.
  
  \begin{enumerate}
     \item {\em Circumstellar zones:} Use table~\ref{tab:rs} to
       interpolate the $r_{is}$ radii for each star using $\Ib = \Ii$.

     \item {\em Circumbinary zone:} Use equation~(\ref{eq:rob}) to get
       the outermost radius $r_{ob}$. For the smallest radii, add to
       the $r_{is}$ previously computed for each star their respective
       star to barycenter distances, the largest of the two is the
       radius of the innermost safe circumbinary edge.

  \end{enumerate}

  \item {\bf Configuration III}: One radiative circumbinary habitable zone whose
    inner edge surrounds both stars. For the outer boundary, use
    equation~(\ref{eq:rob}) to compute $r_{ob}$. For the inner
    boundary, use table~\ref{tab:rib} to compute the distances of each
    star to its closest $\eta_x$ intersection of the inner boundary
    isopleth $r_\star$. To each add the barycenter to respective star
    distance using equation~(\ref{eq:qshift}). The largest of the two
    is $r_{ib}$.

\end{enumerate}

\begin{figure}
\begin{center}
\includegraphics[width=\linewidth]{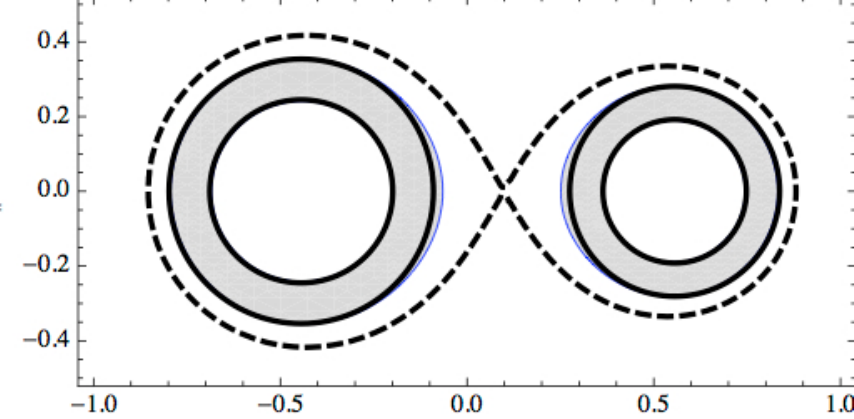}
\caption{The critical flux isoplet (thick, dashed line) and radiative binary
  habitability zones (shaded areas) for the first worked example. The
  limiting safe orbits are shown (thick, continuous lines). The frame
  is centered in the barycenter and the unit of length is the binary
  separation, which in this case is 4 $AU$.}
\label{fig:example1}
\end{center}
\end{figure}

Figure~(\ref{fig:example1}) shows the result for a specific
example. We have a primary G2V star with 1$M_\odot$ and 1$L_\odot$,
while the secondary is a G8V star with 0.8$M_\odot$ and
0.6$L_\odot$. The stars are assumed to be 4$AU$ apart. We have used
the stellar flux limits of \cite{Kopparapu2013} for a G2 star:
$I_o=0.53$ and $I_i=1.10\,L_\odot/AU^2$, which correspond to the
$CO_2$ condensation and water loss limits, respectively. In this case
it is clear that safe orbits can exist within two circumstellar zones.

\begin{figure}
\begin{center}
\includegraphics[width=\linewidth]{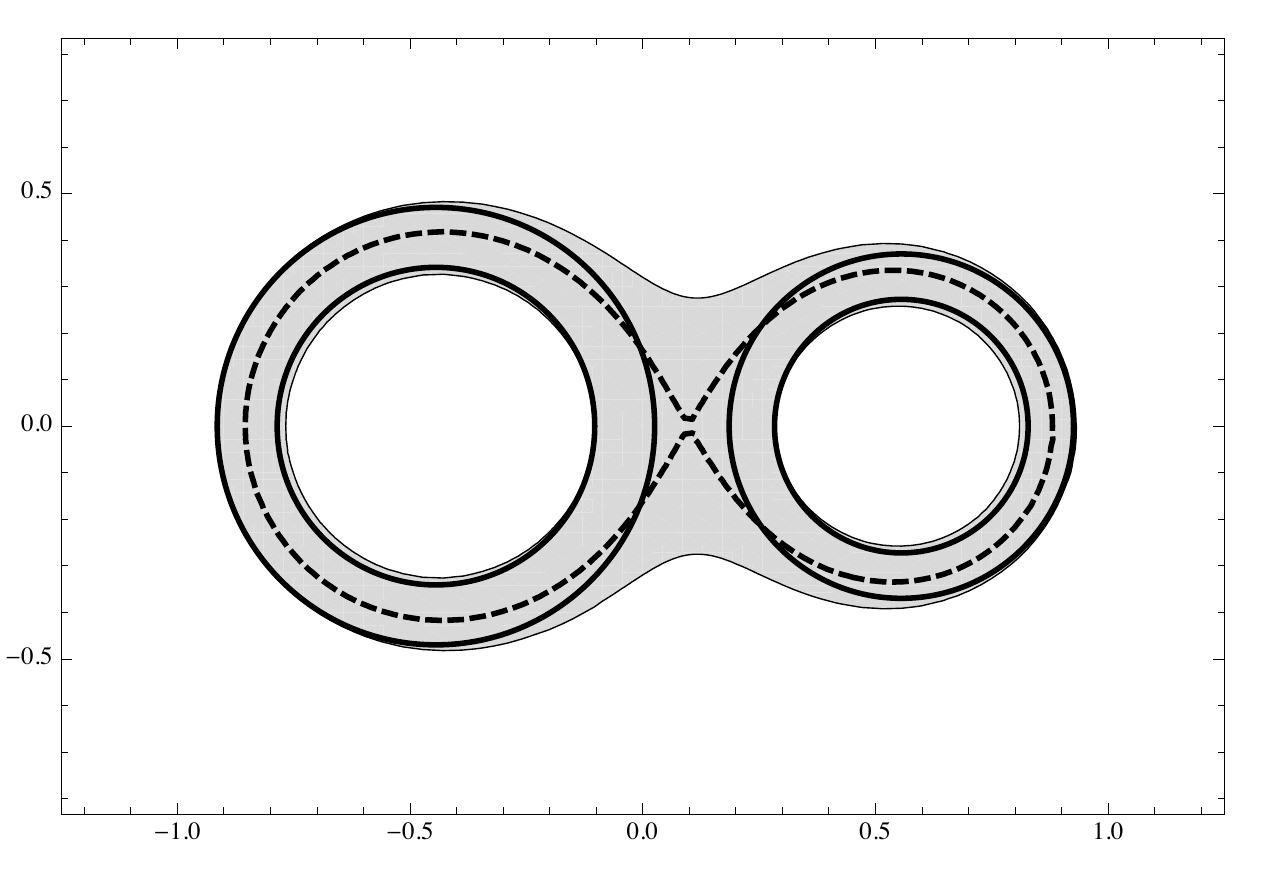}
\caption{Same as figure~(\ref{fig:example1}), but in this case the
  binary separation has shrunk to 3 $AU$ (taken as unit length in the
  plot). Notice that the largest circumstellar orbits reach outside
  the critical isopleth.}
\label{fig:example2}
\end{center}
\end{figure}

Figure~(\ref{fig:example2}) shows the result for the same binary, but
now with a separation of 3$AU$. By putting the stars closer together,
the circumstellar safe zones have merged into a single, case~II
habitable zone. In this case, the only safe orbits lie between the
$r_{is}$ and the outer boundary isopleth. No circumbinary safe orbits
are possible.

\subsection{Binaries in eccentric orbits}

Once the case of the binary in a circular orbit has been solved, the
eccentric orbit case is rather simple. We use the circular orbit
procedure twice: at periastron and apoastron, and compare the
resulting limits. For the smallest safe orbit, we take the largest of
the computed values for both orbital extrema. For the largest safe
orbit, we take the smallest of the respective values. This guarantees
that a fixed, circular, planetary orbit will remain within the safe
zone for all orbital phases. Again, if the resulting outer limit is
smaller than the inner limit, no safe orbit is possible.

We must stress again that, although this procedure does not take into
account explicitly non--circular orbits for planets, nor does it take
into account the possible variation in their shape as a function of
binary orbital phase (if any), we claim that planetary orbits in
binary eccentric systems will not develop readily, highly eccentric
stable orbits ({\it i.e.} $e\la 0.3$). Furthermore, these deforming
effects on orbits in binary systems, may arise close to the critical
isopleth, however, it does not necessarily coincides with the loop gap
between circumstellar and circumbinary stable orbits.

\section{Stability condition for habitable zones in binary stars}
\label{Sec:StabCrit}

In this paper we have employed the criteria of \cite{PSA1} and
\cite{PSA2} for circumestellar and circumbinary stable orbits
respectively. The same approach was used in \cite{JPA}, where
stability zones were studied for binaries of the solar
neighborhood. We show in this paper why this stability method results
better in searching habitable regions for planets.

Unlike the fundamental work of \cite{HW99}, that represents an
excellent empirical approximation to the stability problem, invariant
loops provide an exact solution for stable orbits in ideal (isolated)
binary systems at any binary eccentricity. In this manner, rather than
orbits that keep stable for some small fraction of the binary star
life, to ensure life development, much more time than a small fraction
of the binary life is needed. With invariant loops, numerical
integrations are used to identify them in phase space, their existence
and moreover, their stability, are supported by integrals of motion in
the extended phase space. As long as the orbital parameters of the
binary are not changed, the stability of the loops is guaranteed
without having further numerical integrations (see Pichardo et
al. 2005, Section 2). Consequently, due to the methodology, Holman \&
Wiegert´s approximation overestimates the available stable regions
since their stability criteria depends on a given time of integration
that the particles keep in orbit (about $10^4$ periods), which results
in a very relaxed criteria when looking for strict stability, needed
for life emergency and its development. Furthermore, invariant loops
show in the circumbinary discs cases, that there is a shift of the
stable region that can be considerable for high eccentricities,
compromising the intersection between stable regions for planets and
habitable regions (as is the case, for example of Kepler 16). The
shift can not be detected with the Holman \& Wiegert method, this was
detected theoretically and presented in Pichardo et al. 2008 paper. We
consider and computed this shift in the present work.

In our approach however, we do not consider multiple stars, nor stars
that have evolved away from the main sequence, or planets in highly
eccentric orbits. In the case of multiple stars, this is first because
it is not straightforward to extend the loop formalism to that case,
and furthermore, multiple systems result generally in unstable
systems, unless they are hierarchical (v.g. triple systems where a
very close binary star with a far companion as the Alpha Centauri
system and Proxima Centauri seem to be, or a system with a double
binary where both binary systems act like a whole binary system since
they tend to be extremely separated), and in that case, these systems
can be reduced to regular separated binary systems at a good
approximation. On the same direction, in the general case of multiple
stars, stable regions for planetary orbits would be severely
compromised, and even if stable regions would exist, finding them
would be a challenging task. Thus, if we are not able to calculate
stable regions, it is not of importance to find the correct irradiance
zones because we still would not be able to establish habitable zones
(i.e. the combination of stability and stellar irradiance).

Regarding the type of star, although our method allows us to calculate
radiative safe zones for any type of star, in our specific study for
solar neighborhood binaries, we only calculated habitable zones for
main sequence stars since those out of this ``life stage'', would
likely be inhospitable for life.

Finally, we do not include studies of planets on very eccentric orbits
(i.e. $e_p\ga 0.3$), in our experience and work with stability zones
in binaries constructed out of invariant loops, we have found that
their probability of survival is highest, for example, either on
planets in a circumbinary disc for the case of very close binary
systems (where the circumbinary material feels the system almost as a
single star), or in circumstellar discs, for very open binaries (where
both stars act almost as single stars). However, this is not the
general case, in any other binary system in between, stable regions
for planetary orbits with high eccentricities ($e_p\ga 0.3$), the
reduction of the available phase space for planets to settle down, and
consequently, of their possibilities to be stable for long timescales
can be considerable (this is an ongoing study that will be presented
in a future paper).

Under this approach stable zones are defined by invariant loops (see
\cite{PSA1} and \cite{PSA2} for details). The radius for the most
exterior dynamically stable orbit around each star is given by:

\begin{equation}
\label{estabilidadCE}
R_{i}=R_{i}^{Egg}\left(0.733(1-e)^{1.2}q^{0.07} \right),
\end{equation}
In a similar manner, the inner viable radius for circumbinary stable
orbits is:

\begin{equation}
\label{estabilidadCB}
R_{CB} \approx 1.93 a \left( 1+1.01e^{0.32} \right)\left( q(1-q)
\right) ^{0.043},
\end{equation}
where subindex $i$ represents each star primary ($i=1$) and secondary
($i=2$), $a$ is the semimajor axis of the binary, $e$ the eccentricity
and $q=q_{2}/(q_{1}+q_{2})$ is the mass ratio (see
eq. (\ref{eq:qshift})).

Finally, in equation (\ref{estabilidadCE}), $R_{i}^{Egg}$ is the
approximation of Eggleton \cite{E83} to the maximum radius of a circle
within the Roche lobe, given by,

\begin{equation}
R_{i}^{Egg}= \frac{ 0.49 a q_{i}^{2/3}}{0.6q_{i}^{2/3}+ln(1+q_{i}^{1/3})},
\end{equation}
in this equation $q_{i}$ is defined by $q_{1}=M_{*1}/M_{*2}$ and
$q_{2}=M_{*2}/M_{*1}$.

For the general eccentric case, one must consider a shift of the
center of the minimum radius for stable circumbinary orbits, $R_{CB}$
( eq. (\ref{estabilidadCB})). This shift is given by,
\begin{equation}
\label{shift}
R_{sh}=-3.7 \ a \ e^{0.8}\left( 0.5-q \right) \left[q(1-q) \right]^{1/4}.
\end{equation}

It is important to stress that the orbits defined by equations
(\ref{estabilidadCE}) and (\ref{estabilidadCB}), represent stable
orbits formed by non self-intersecting loops, where gas and planets
could settle down in long term basis.

\section{The combined habitability condition: some particular cases.}
\label{Sec:partcases}

As we have mentioned, the combination of the radiative and stability 
conditions is one of the goals of the present paper. In order
to show how this approach allows us to find candidates in the search
for habitable planets, in this section we apply our method to 64 binary systems 
of the solar neighborhood, with known orbital parameters, and main sequence stars.
Most of the cases are eccentric
binaries, thus we apply our procedure (both, radiative and stability conditions) 
at periastron and apoastron. Once
we have these two calculations, we compare both and take,
conservatively, the most restricted circular edge. It is worth to
mention that it is important to compute it in both locations because
we do not know {\it a priori} in which case the binary will be located
({\it i.e.}, if the habitable zone will be circumstellar or
circumbinary). Given this, the restriction at periastron or apoastron
acts in different ways for configurations I, II or III, and the case can even
change from one to another, and this should be taken into account.

Most of binaries considered in this paper are taken from \cite{JPA},
which correspond to binaries with stars in the main sequence. We have
applied the classical stellar luminosity-mass function \cite{COX} in
order to obtain the luminosity for each star. Orbital parameters are
the same as the ones considered in \cite{JPA}.

Another important issue that must be considered, is the circumbinary
disc center shift, produced by eccentric binary systems
(eq.(\ref{shift})). In high eccentric cases this shift might become
relevant when calculating the circumbinary orbits inside the habitable
zone.

Table 3 shows the results for the 64 objects at periastron, column $1$
is the Hipparcos name, column $2$ is an alternative name, column $3$
shows the particular habitable case resulting for the binary, columns
$4$ and $5$ contain the inner and outer habitable radii for the
primary star and columns $6$ and $7$ are the same but around the
secondary star. Finally columns $8$ and $9$ provide the inner and
outer habitable radii in the circumbinary case. Table 4 shows the same
as Table 3 but at apoastron.

Table 5 shows orbital parameters for each binary, columns $1$ and $2$
are the HIP name and alternative name, column $3$, $4$, $5$ and $6$
are semimajor axis, eccentricity, primary and secondary masses
respectively. Column $7$ shows the outer stability radius around the
primary star while column $8$ shows the same but for the secondary
star of the binary. Column $9$ shows the inner circumbinary stability
radius for the object and column $10$ gives the shift for the
circumbinary stability radius. Finally column $11$ provides the
reference for orbital parameters used in this paper.

Table 6 shows the cases where all criteria are satisfied together,
{\it i.e.} objects where the intersection of habitability at
periastron and apoastron is not empty. For the objects in this table
we can observe some habitability zone well defined at periastron and
apoastron at the same time and also inside of the stability zone given
by equations of section \ref{Sec:StabCrit}. In circumbinary cases the
shift was taken into account.

In this section three particular cases are considered in order to show
how the approach is implemented.

\begin{figure}
\begin{center}
\includegraphics[width=\linewidth]{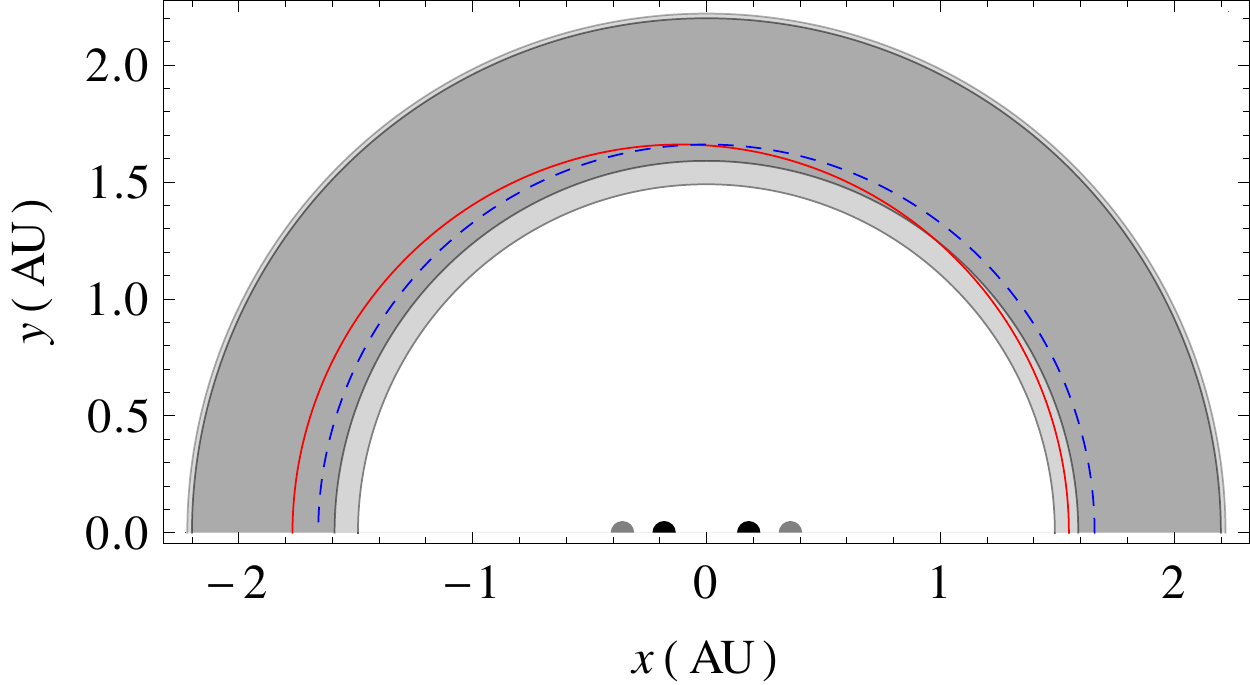}
\caption{Effective habitable zone (radiative and stability conditions)
  for HIP 1955. Black dots are the stars at periastron, gray dots at
  apoastron. The shaded region is the radiative habitable zone, with
  darker gray indicating the intersection between these two zones at
  periastron and apoastron. The blue dashed circle represents the
  minimum dynamical stable orbit around the binary and the red one is
  the same but considering the effect because the shift.}
\label{fig:1955}
\end{center}
\end{figure}

{\it HIP 1995} \\ HIP 1995 has a semimajor axis of $a=0.54 AU$ with an
eccentricity $e=0.33$, stellar masses are $M_{*1}=1.13M_{\odot}$ and
$M_{*2}=0.45M_{\odot}$. We have included habitable zones at periastron
and apoastron (tables 3 and 4), this two separate calculations will
restrict the effective habitable zone, which is located just in a
circumbinary position. At Periastron we have $R^{*1}_{P}(inner)=1.49
AU$, $R^{*1}_{P}(outer)=2.22$, and at apoastron
$R^{*1}_{A}(inner)=1.59 AU$, $R^{*1}_{A}(outer)=2.2$. The dynamical
stability is given by the minimum radius of the circumbinary stable
zone, following equation (\ref{estabilidadCB}) this starts at
$R_{CB}=1.66 AU$. As an additional restriction, the shift should be
considered, its value in this case is important because it change the
viable zone defined under our approach, $R_{shift}=-0.11AU$. Figure
\ref{fig:1955} shows all of these estimations, black (semi) dots are
the stars located at periastron while (semi) dots in gray show them at
apoastron. Gray disks are the habitable zones, calculated for both
cases. Dark gray provides the intersection of the BHZ, the effective
habitable zone for the binary. Dashed blue circle is the minimum
dynamically stable circumbinary orbit, and the red one is the same but
with the shift considered, this is the inner border of the viable
stable orbit. Taken in to account all of these criteria is how we
decide where is located the completely viable BHZ.

\begin{figure}
\includegraphics[width=\linewidth]{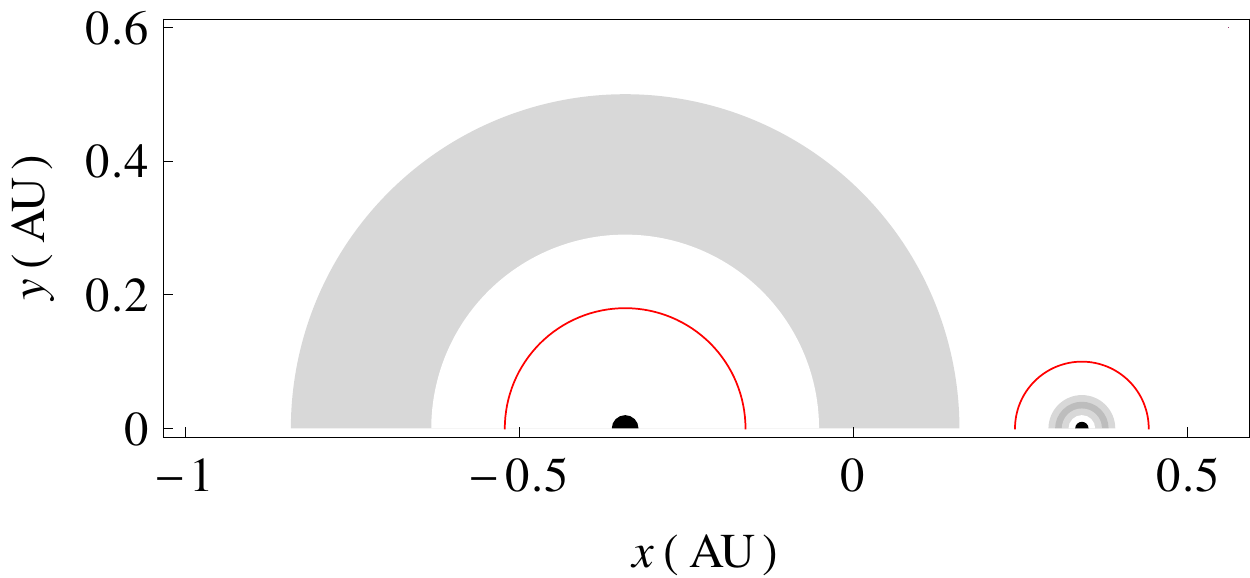}
\caption{Effective habitable zone for HIP 80346. Black dots are the
  stars located at periastron, gray disks shows the habitable zone,
  darker gray disk illustrate the intersection of habitable at
  periastron and apoastron. Red circles shows the maximum dynamical
  stable orbit around each star. We can observe only secondary star
  have a complete effective habitable zone.}
\label{Fig:HIP80346ab}
\end{figure}

\begin{figure}
\includegraphics[width=\linewidth]{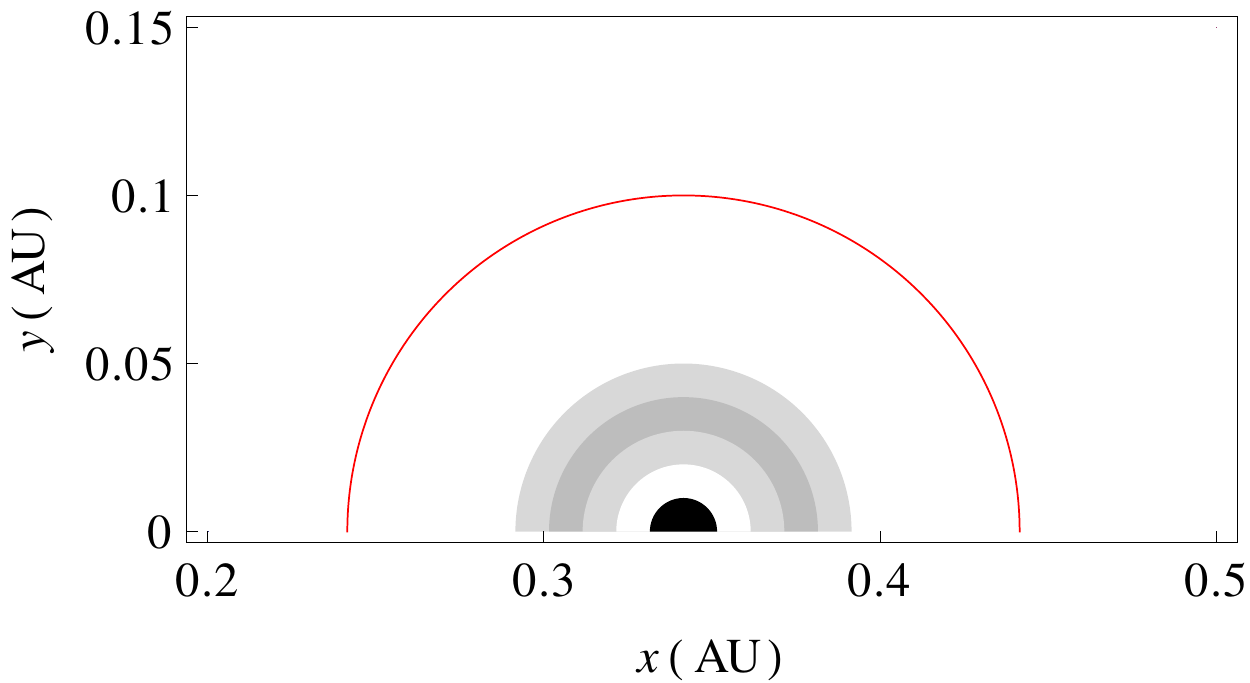}
\caption{Zoom of secondary star of HIP 80346. This star can provide a
  complete effective habitable zone. Darker gray disk is where we can
  expect an habitable planet.}
\label{Fig:HIP80346b}
\end{figure}

{\it HIP 80346} \\ This is a particular case, in this object we can
found an effective habitability within the dynamical stability zone
just around the secondary star. Semimajor axes of this binary is
$a=2.07 AU$, eccentricity $e=0.67$, nevertheless stellar masses are
very small, $M_{*1}=0.5M_{\odot}$ and $M_{*2}=0.13$. The main
restriction to habitability is because the eccentricity of the
binary. The dynamical stability region is given by equation
\ref{estabilidadCE}, by using this relation we obtain the maximum
stable orbit radius around each star, $R_{ce}^{*1}=0.18 AU$ and
$R_{ce}^{*2}=0.1 AU$. Figure \ref{Fig:HIP80346ab} shows both stars at
periastron, the gray disks are the habitable zones, where the darker
gray allows to see the habitable zone with the value at apoastron
considered. Red circle around each star provides the maximum radius
for dynamical stable orbits. We can observe that for the primary star
habitable zone is out of dynamical stability, but in the secondary we
can observe habitability within stable zone. Figure
\ref{Fig:HIP80346b} shows a zoom for the secondary star of this
object.

{\it HIP 76852} 
\\ 

\begin{figure}
\includegraphics[width=\linewidth]{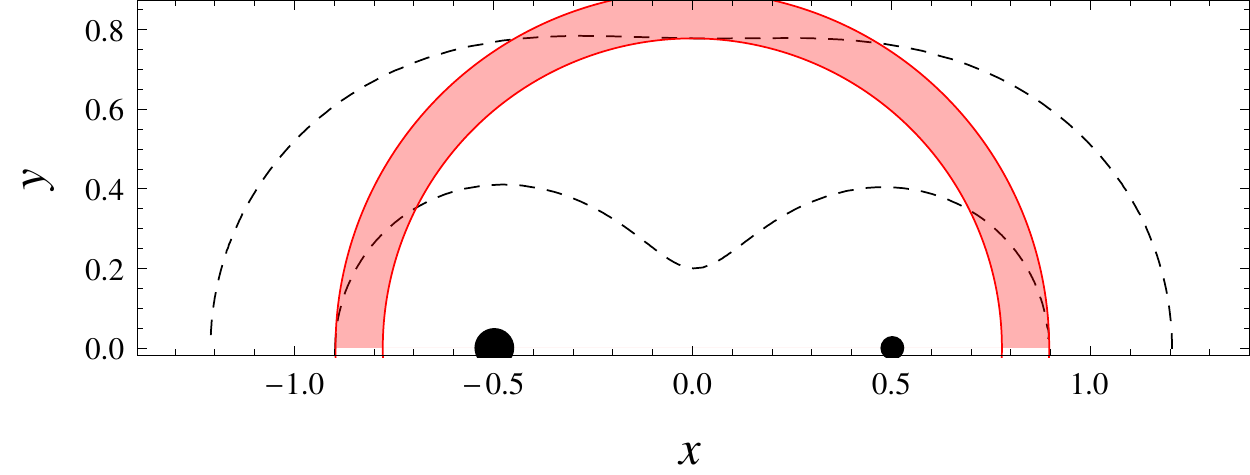}
\caption{A pathological case, HIP 76852: Dashed lines show the boundaries of the radiative
       safe zone. This is a type III configuration. However, as shown by the red circles,
       no circumbinary BHZ is possible despite having a non-zero radiative safe zone.}
\label{Fig:raro}
\end{figure}

HIP 76852 is a peculiar configuration III case, because we can not
find a well defined BHZ (the inner boundary has a radius larger than
the outer one), and yet, it has a non-zero radiative safe zone. Figure
\ref{Fig:raro} shows this case (in units of the separation at
periastron), black dots are primary (left) and secondary (right) stars
located at periastron, interior dashed line is the flux isoplete that
define the inner habitable zone, the exterior dashed line is the same
but for the outer boundary. The inner red circle is what we have
defined as the minimum circunscrite circle to the outer habitable
boundary and the exterior red circle is the circle that circunscribe
the inner flux isoplete boundary for habitability. This way we can
observe it is not possible to have both conditions satisfied at the
same time, nevertheless an habitable zone is well defined around the
binary. This is a very important restriction because, in case we have
this kind of objects we can not tell for sure if a planet can be
hosted within this zone, the eccentricity that the planet must have in
order to be all the time inside the habitable zone could be very high
or even the shape of the orbit could be non physical. So this case
show us it is not always possible to find a well behaved habitable
zone, where some planet can settle.

\section{Discussion and Conclusions}
\label{Sec:conclusions}

In this work we have constructed a straightforward and clear
formulation to calculate regions for habitable planets in binary
stellar systems. To this purpose, we search for two general
restrictions assuming in principle Earth-like planets: a) the planet
must be located in a region of orbital stability (and approximately
circular orbit), and b) the planet is located at a position, such as
it reaches the correct host star energy, to permit the existence of
liquid water on its surface \cite{Kopparapu2013}. Some other
particular restrictions, as the ones proper of the intrinsic
characteristics of the planet for example, can readily be addressed to
this formulation as multiplicative factors.

Regarding the fundamental test for orbital stability to develop life,
in this paper we employ the criteria of \cite{PSA1} and \cite{PSA2}
for stable orbits in binary systems. This stability method results
better in searching habitable regions for planets. Indeed, although
the empirical approximation of Holman \& Wiegert was an excellent and
fundamental first approximation to the stability problem, their method
is rather based on a detailed trial and error approximation, that
overestimates by construction, the available stable regions since
their stability criteria depends on a given time of integration that
particles keep in orbit (about $10^4$ periods), which results on a
relaxed criteria when looking for strict stability, needed for life
emergency and developing. Instead, we have employed the invariant
loops method that searches for the exact solution for stable orbits in
binary systems at all binary eccentricities.

On the other hand, invariant loops show in the circumbinary discs
cases, that there is a shift of the geometric center of the stable
region (disk) that can be considerable for high eccentricities,
compromising the intersection between stable regions for planets and
habitable regions in the simple sense of the irradiance (as is the
case, for example of Kepler 16). The shift can not be detected with
other method than the exact solution provided by the invariant loops
tool, this was detected theoretically and presented in Pichardo et
al. 2008 paper, we are considering this for planets in circumbinary
disks in our calculations for binaries of the solar neighborhood.

It is worth mentioning that in our approach to calculate habitable
zones in binary stars, for different life likelihood considerations,
such as orbital stability and irradiation, we are not considering
multiple stars, or stars out of their main sequence or planets on
highly eccentric orbits. In the case of multiple stars, although an
extension of this work to any multiple star systems would be
straightforward for the irradiance calculations, in the majority of
cases these systems result in unstable ones, unless they are
hierarchical (v.g. triple systems where a very close binary star with
a far companion as Alpha Centauri and Proxima, or a double binary
where both binary systems act like single stars in one larger binary
system). On the other hand, there is no straightforward method to find
stable orbital zones in multiple stellar systems, needed to calculate
formally habitable zones.

Although our method is aplicable to all kind of stars, we only
calculated habitable zones for the solar neighborhood main sequence
stars, since stars out of this life stage would likely be inhospitable
for life. In the case of planets on very eccentric orbits (e$\ga$0.3),
the probability of survival is diminished due to the presence of the
companion that reduces severely the available phase space for planets
to settle down (this is a detailed study that will be presented in a
future work). On the other hand, what induces eccentricity on a planet
in a binary star?, external factors are stellar encounters, but to
affect a very small disc (truncated by the binary), a very close
stellar encounter (with a third star) must be taking place ($\sim$3
times the disk radius at least, \citet{JT11}), likely affecting the
binary stability itself, this results in a non stable situation for
habitability. Or an internal factor, for example a resonance that is
able to induce secularly eccentricity in the planetary orbits, however
that is also an unstable situation for planets, since resonances tend
to increase rapidly eccentricities on orbiting bodies, wiping out
entire disk regions. Other possibility, a giant planet on eccentric
orbit that induces eccentricity in terrestrial planets, a clearly
difficult situation in terms of stability.

Regarding irradiance zones. We have defined three zones where the
binary star provides the necessary energy for habitability: (I) the
zone around each star, (III) the zone around the hole binary or (II)
in a mixture of this two zones. In this work, we consider a
``habitable environment'', the intersection of one of these zones and
the allowed dynamical zone for stable orbits.

Taking into account both restrictions, from a binary sample of main
sequence stars, with known orbital parameters of the solar
neighborhood (64), we have selected 36 candidates (56 \% of our
original sample), as plausible candidates to host habitable planets.
We present this table together with three particular and interesting
examples in detail: HIP 1995, HIP 80346 and HIP 76852.

We find, from our sample of candidates, that none allows planets
inside the BHZ defined by the configuration II (circumbinary discs). This is
because in all cases, the system allows habitability too close to the
binary, where the stability restriction becomes very important, making
impossible for all cases in our sample, to host a planet
there. Although a greater sample is necessary to produce a final
conclusion on the solar neighborhood, this small first sample is
useful to statistically elucidate the possibilities of finding
habitability on binaries of the solar neighborhood, and the
possibilities for circumbinary discs seem reduced.

Programming our formulation for habitable zones (from the irradiance
point o view) is rather simple, however software to compute numerically
the size of habitable zones for binary systems is available from the
authors.

\begin{table*}
\label{tabla-peri}
\begin{tabular}{cccccccccccccccc}
\label{tabla-peri}

%\hline

 HIP	 & Alter. name	& Case$_{P}$	&R$^{*1}_{P}$ inner	&R$^{*1}_{P}$ outer	&R$^{*2}_{P}$ inner	&R$^{*2}_{P}$ outer	&R$^{**}_{P}$ inner	&	R$^{**}_{P}$ outer	\\

 	&		&		&	[AU]			&[AU]			&[AU]			&[AU]			&[AU]			&	[AU]			\\
\hline

--			&			$\delta$ Equ 		&		3		&		-		&		-		&		-		&		-		&		4.51		&		6.49	\\
--			&			V821 Cas		&		3		&		-		&		-		&		-		&		-		&		5.04		&		8.55	\\
1349			&			HD 1273			&		3		&		-		&		-		&		-		&		-		&		1.26		&		1.78	\\
1955			&			HD 2070			&		3		&		-		&		-		&		-		&		-		&		1.49		&		2.22	\\
2941			&			ADS520 			&		1		&		0.55		&		0.95		&		0.55		&		0.95		&		-		&		-	\\
5842			&			HD 7693 		&		1		&		0.87		&		1.50		&		0.78		&		1.34		&		-		&		-	\\
7078			&			HD 9021			&		3		&		-		&		-		&		-		&		-		&		1.78		&		2.70	\\
7751			&			HD 10360		&		1		&		0.66		&		1.14		&		0.63		&		1.08		&		-		&		-	\\
7918			&			HD 10307		&		1		&		0.71		&		1.22		&		0.03		&		0.05		&		-		&		-	\\
8903			&			HD 11636		&		3		&		-		&		-		&		-		&		-		&		4.68		&		7.68	\\
11231			&			HD 15064		&		3		&		-		&		-		&		-		&		-		&		1.35		&		2.02	\\
12062			&			HD 15862		&		3		&		-		&		-		&		-		&		-		&	 - 	&	 -\\
12153			&			HD 16234		&		3		&		-		&		-		&		-		&		-		&		128.82		&		219.85	\\
12623			&			HD 16739		&		3		&		-		&		-		&		-		&		-		&		2.53		&		4.16	\\
\bf{14954}		&			\bf{HD 19994}		&		1		&		1.92		&		3.30		&		0.15		&		0.25		&		-		&		-	\\
18512			&			HD 24916		&		1		&		0.15		&		0.25		&		0.04		&		0.06		&		-		&		-	\\
20087			&			HD 27176		&		2		&		3.50		&		4.44		&		1.41		&		2.28		&	- 	&	-\\
20935			&			HD 28394		&		3		&		-		&		-		&		-		&		-		&		2.02		&		3.25	\\
24419			&			HD 34101		&		2		&		0.89		&		1.53		&		0.07		&		0.18		&		1.37		&		1.27$^*$\\
30920			&				&		1		&		0.06		&		0.10		&		0.01		&		0.02		&	- 	&	-\\
33451			&			HD 51825		&		3		&		-		&		-		&		-		&		-		&	- 	&	-\\
34164			&			HD 53424		&		3		&		-		&		-		&		-		&		-		&		1.77		&		2.17	\\
39893			&						&		3		&		-		&		-		&		-		&		-		&		1.50		&		1.53	\\
44248			&				&		1		&		2.19		&		3.75		&		0.90		&		1.60		&	- 	&	-\\
45343			&				&		1		&		0.31		&		0.54		&		0.30		&		0.52		&	- 	&	-\\
56809			&			HD 101177		&		1		&		3.86		&		6.63		&		1.95		&		3.34		&		-		&		-	\\
63406			&			HD 112914		&		2		&		0.76		&		0.96		&		0.11		&		0.30		&		0.98		&		1.10	\\
64241			&				&		2		&		1.90		&		3.17		&		1.47		&		2.48		&	- 	&	-\\
64797			&				&		1		&		0.60		&		1.03		&		0.31		&		0.54		&	- 	&	-\\
\bf{67275}		&			\bf{HD 120136}		&		1		&		1.92		&		3.30		&		0.19		&		0.33		&		0		&		-	\\
67422			&				&		1		&		0.58		&		1.00		&		0.48		&		0.82		&	- 	&	-\\
72848			&			HD 131511   		&		3		&		-		&		-		&		-		&		-		&		1.05		&		1.56	\\
73440			&			HD 133621		&		3		&		-		&		-		&		-		&		-		&		1.27		&		1.83	\\
75379			&			HD 137502		&		3		&		-		&		-		&		-		&		-		&		1.84		&		2.86	\\
76852			&			HD 140159		&		3		&		-		&		-		&		-		&		-		&	- 	&	-\\
79101			&			HD 145389		&		3		&		-		&		-		&		-		&		-		&		11.98		&		18.51	\\
80346			&						&		1		&		0.29		&		0.50		&		0.03		&		0.05		&		-		&		-	\\
80686			&			HD 147584		&		3		&		-		&		-		&		-		&		-		&		1.23		&		1.90	\\
82817			&			HD 152771		&		1		&		0.36		&		0.62		&		0.14		&		0.25		&		-		&		-	\\
82860			&			HD 153597		&		3		&		-		&		-		&		-		&		-		&		1.59		&		2.42	\\
84425			&				&		2		&		1.73		&		2.82		&		0.97		&		1.61		&	- 	&	-\\
84720			&			HD 156274		&		1		&		0.69		&		1.19		&		0.26		&		0.44		&		-		&		-	\\
84949			&			HD 157482		&		3		&		-		&		-		&		-		&		-		&		7.35		&		11.01	\\
86400			&			HD 1360346		&		3		&		-		&		-		&		-		&		-		&		0.70		&		0.98	\\
86722			&			HD 161198		&		3		&		-		&		-		&		-		&		-		&		1.03		&		1.54	\\
87895			&			HD 163840		&		3		&		-		&		-		&		-		&		-		&		1.61		&		1.88	\\
91768			&			HD 173739		&		1		&		0.18		&		0.31		&		0.14		&		0.24		&		-		&		-	\\
\bf{93017}		&			\bf{ADS 11871}		&		1		&		2.86		&		4.93		&		2.64		&		4.56		&		-		&		-	\\
93825			&				&		1		&		1.72		&		2.95		&		1.66		&		2.87		&	- 	&	-\\
95028			&			HD 181602		&		3		&		-		&		-		&		-		&		-		&		2.21		&		3.28	\\
95575			&			HD 183255		&		3		&		-		&		-		&		-		&		-		&		0.86		&		1.10	\\
\bf{98001}		&			\bf{HD 188753}		&		2		&		1.88		&		3.16		&		1.43		&		2.43		&	- 	&	-\\
99965			&			HD 193216		&		3		&		-		&		-		&		-		&		-		&		1.31		&		1.42	\\
109176			&			HD 210027		&		3		&		-		&		-		&		-		&		-		&		1.83		&		2.99	\\
111170			&			HD 213429		&		3		&		-		&		-		&		-		&		-		&		1.72		&		2.21	\\
113718			&			HD 217580		&		3		&		-		&		-		&		-		&		-		&		0.75		&		1.00	\\
\bf{116310}		&			\bf{HD 221673}		&		1		&		4.06		&		6.99		&		4.06		&		6.99		&		-		&		-	\\
\bf{116727}		&			\bf{HD 222404}		&		1		&		2.62		&		4.50		&		0.20		&		0.35		&		-		&		-	\\
\bf{}			&			\bf{Kepler 16}		&		3		&		-		&		-		&		-		&		-		&		0.58		&		0.85	\\
\bf{}			&			\bf{Kepler 34}		&		3		&		-		&		-		&		-		&		-		&		1.65	&		2.81	\\
				&		Kepler 35		&		3		&	-	&	-	&	-	&	-	&		1.76		&		2.75	\\
				&		Kepler 38		&		3		&	-	&	-	&	-	&	-	&		0.64		&		0.99	\\
				&		Kepler 47		&		3		&	-	&	-	&	-	&	-	&		0.12		&		0.21	\\
				&		Kepler 64		&		3		&	-	&	-	&	-	&	-	&		2.60		&		4.44	\\

\hline
\end{tabular}
\caption{Radiative habitable zones at Periastron. Columns are: (1) HIP name, (2) Alternative name (some cases), 
(3) rHZ case described in section \ref{HZ}, (4) Inner rHZ for primary star, (5) Outer rHZ for primary star, 
(6) Inner rHZ for secondary star, (7) Outer rHZ for secondary star, (8) Inner rHZ for binary zone and (9) 
Outer rHZ for binary zone. Note: rHZ - Radiative Habitable Zone.}
\end{table*}

\begin{table*}
\label{tabla-apo}
\begin{tabular}{ccccccccccccccc}

 HIP 	& Alter. name	&	 Case$_{A}$	& R$^{*1}_{A}$ inner	& R$^{*1}_{A}$ outer	& R$^{*2}_{A}$ inner	& R$^{*2}_{A}$ outer	& R$^{**}_{A}$ inner	& R$^{**}_{A}$ outer	\\

 	&	&		&[AU]			&[AU]			&[AU]			&[AU]			&[AU]			&	[AU]			\\

\hline

--			&			$\delta$ Equ 		&		3		&		-		&		-		&		-		&		-		&	-	&	-\\
--			&			V821 Cas		&		3		&		-		&		-		&		-		&		-		&		5.04		&		8.55	\\
1349			&			HD 1273			&		3		&		1.16		&		1.82		&		0.50		&		0.79		&	-	&	-\\
1955			&			HD 2070			&		3		&		-		&		-		&		-		&		-		&		1.59		&		2.20	\\
2941			&			ADS520 	 		&		1		&		0.55		&		0.95		&		0.55		&		0.95		&		-		&		-	\\
5842			&			HD 7693  		&		1		&		0.87		&		1.50		&		0.78		&		1.34		&		-		&		-	\\
7078			&			HD 9021			&		3		&		-		&		-		&		-		&		-		&		1.91		&		2.67	\\
7751			&			HD 10360 		&		1		&		0.66		&		1.13		&		0.63		&		1.08		&		-		&		-	\\
7918			&			HD 10307		&		1		&		0.71		&		1.22		&		0.02		&		0.04		&		-		&		-	\\
8903			&			HD 11636		&		3		&		-		&		-		&		-		&		-		&		5.01		&		7.66	\\
11231			&			HD 15064		&		3		&		-		&		-		&		-		&		-		&		1.48		&		2.00	\\
12062			&			HD 15862		&		3		&		1.00		&		1.70		&		0.25		&		0.47		&	-	&	-\\
12153			&			HD 16234		&		3		&		-		&		-		&		-		&		-		&		129.76		&		219.82	\\
12623			&			HD 16739		&		3		&		-		&		-		&		-		&		-		&		3.09		&		4.03	\\
\bf{14954}		&			\bf{HD 19994}		&		1		&		1.92		&		3.30		&		0.15		&		0.26		&		-		&		-	\\
18512			&			HD 24916		&		1		&		0.15		&		0.25		&		0.04		&		0.06		&		-		&		-	\\
20087			&			HD 27176		&		2		&		3.24		&		5.50		&		1.10		&		2.00		&	-	&	-\\
20935			&			HD 28394		&		3		&		-		&		-		&		-		&		-		&		2.17		&		3.21	\\
24419			&			HD 34101		&		2		&		0.89		&		1.53		&		0.06		&		0.15		&	-	&	-\\
30920		&				&		1		&		0.06		&		0.10		&		0.01		&		0.02		&	-	&	-\\
33451			&			HD 51825		&		3		&		2.72		&		4.67		&		1.73		&		3.02		&		-		&		-	\\
34164			&			HD 53424		&		3		&		-		&		-		&		-		&		-		&	-	&	-\\
39893			&						&		3		&		1.02		&		1.71		&		0.37		&		0.67		&	-	&	-\\
44248		&				&		1		&		2.18		&		3.74		&		0.89		&		1.55		&	-	&	-\\
45343		&				&		1		&		0.31		&		0.54		&		0.30		&		0.52		&	-	&	-\\
56809			&			HD 101177		&		1		&		3.86		&		6.63		&		1.95		&		3.35		&		-		&		-	\\
63406			&			HD 112914		&		2		&		0.75		&		1.28		&		0.07		&		0.14		&		-		&		-	\\
64241		&				&		1		&		1.79		&		3.09		&		1.35		&		2.34		&	-	&	-\\
64797		&				&		1		&		0.60		&		1.03		&		0.31		&		0.54		&	-	&	-\\
\bf{67275}		&			\bf{HD 120136}		&		1		&		1.92		&		3.30		&		0.19		&		0.33		&		-		&		-	\\
67422		&				&		1		&		0.58		&		1.00		&		0.48		&		0.82		&	-	&	-\\
72848			&			HD 131511   		&		3		&		-		&		-		&		-		&		-		&		1.21		&		1.52	\\
73440			&			HD 133621		&		3		&		1.15		&		1.97		&		0.06		&		0.46		&		1.38		&		1.79	\\
75379			&			HD 137502		&		3		&		-		&		-		&		-		&		-		&		2.24		&		2.77	\\
76852			&			HD 140159		&		3		&		4.43		&		7.34		&		4.35		&		7.21		&	-	&	-\\
79101			&			HD 145389		&		3		&		-		&		-		&		-		&		-		&		12.53		&		18.46	\\
80346			&						&		1		&		0.29		&		0.50		&		0.02		&		0.04		&		-		&		-	\\
80686			&			HD 147584		&		3		&		-		&		-		&		-		&		-		&		1.23		&		1.90	\\
82817			&			HD 152771		&		1		&		0.36		&		0.62		&		0.14		&		0.24		&		-		&		-	\\
82860			&			HD 153597		&		3		&		-		&		-		&		-		&		-		&		1.63		&		2.42	\\
84425		&				&		1		&		1.61		&		2.78		&		0.82		&		1.43		&	-	&	-\\
84720			&			HD 156274		&		1		&		0.69		&		1.19		&		0.26		&		0.44		&		-		&		-	\\
84949			&			HD 157482		&		3		&		-		&		-		&		-		&		-		&		9.28		&		10.38	\\
86400			&			HD 1360346		&		3		&		-		&		-		&		-		&		-		&		0.76		&		0.96	\\
86722			&			HD 161198		&		3		&		0.97		&		1.66		&		0.14		&		0.25		&		-		&		-	\\
87895			&			HD 163840		&		3		&		1.11		&		1.86		&		0.58		&		1.01		&	-	&	-\\
91768			&			HD 173739		&		1		&		0.18		&		0.31		&		0.14		&		0.24		&		-		&		-	\\
\bf{93017}		&			\bf{ADS 11871}		&		1		&		2.82		&		4.87		&		2.60		&		4.49		&		-		&		-	\\
93825		&				&		1		&		1.71		&		2.94		&		1.66		&		2.85		&	-	&	-\\
95028			&			HD 181602		&		3		&		-		&		-		&		-		&		-		&		2.37		&		3.25	\\
95575			&			HD 183255		&		3		&		-		&		-		&		-		&		-		&		0.92		&		1.07	\\
\bf{98001}		&			\bf{HD 188753}		&		2		&		1.79		&		3.09		&		1.33		&		2.30		&		-		&		-	\\
99965			&			HD 193216		&		3		&		-		&		-		&		-		&		-		&		1.38		&		1.38	\\
109176			&			HD 210027		&		3		&		-		&		-		&		-		&		-		&		1.83		&		2.99	\\
111170			&			HD 213429		&		3		&		-		&		-		&		-		&		-		&	-	&	-\\
113718			&			HD 217580		&		3		&		0.64		&		1.11		&		0.04		&		0.09		&		-		&		-	\\
\bf{116310}		&			\bf{HD 221673}		&		1		&		4.05		&		6.97		&		4.05		&		6.97		&		-		&		-	\\
\bf{116727}		&			\bf{HD 222404}		&		1		&		2.62		&		4.50		&		0.19		&		0.32		&		-		&		-	\\
\bf{}			&			\bf{Kepler 16}		&		3		&		-		&		-		&		-		&		-		&		0.59		&		0.85	\\
\bf{}			&			\bf{Kepler 34}		&		3		&		-		&		-		&		-		&		-		&		1.67		&		8.90	\\
		&		Kepler 35		&		3		&	-	&	-	&	-	&	-	&		2.44		&		3.80	\\
		&		Kepler 38		&		3		&	-	&	-	&	-	&	-	&		0.71		&		1.09	\\
		&		Kepler 47		&		3		&	-	&	-	&	-	&	-	&		0.11		&		0.19	\\
		&		Kepler 64		&		3		&	-	&	-	&	-	&	-	&		1.18		&		6.26	\\

\hline

\end{tabular}
\caption{Radiative habitable zones at Apoastron. Columns are: (1) HIP name, 
(2) Alternative name (some cases), (3) rHZ case described in section \ref{HZ}, (4) Inner rHZ for primary star, 
(5) Outer rHZ for primary star, (6) Inner rHZ for secondary star, (7) Outer rHZ for secondary star, 
(8) Inner rHZ for binary zone and (9) Outer rHZ for binary zone. Note: rHZ - Radiative Habitable Zone.}
\end{table*}

\begin{table*}
\begin{tabular}{ccccccccccccccc}
\label{tabla-estabilidad}

HIP	&	Alter. name	&	a	&	e	&	m$^{*1}$	&	m$^{*2}$	&	r$_{ce}^{*1}$	&	r$_{ce}^{*2}$	&	r$_{**}$	&	r$_{shift}$	&	ref	\\
	&			&	[AU]	&		&	M$_{\odot}$	&	M$_{\odot}$	&	[AU]		&	[AU]		&	[AU]		&	[AU]	&		\\
\hline

--				&			$\delta$ Equ			&			4.73			&			0.42			&			1.66			&			1.59			&			0.66			&			0.64			&			15.18			&			-0.07			&			A	\\
--				&			V821 Cas			&			0.044			&			0.13			&			2.046			&			1.626			&			0.01			&			0.01			&			0.12			&			0.00			&			B	\\
1349				&			HD 1273				&			1.25			&			0.57			&			0.98			&			0.55			&			0.13			&			0.1			&			4.18			&			-0.29			&			C	\\
1955				&			HD 2070				&			0.54			&			0.33			&			1.13			&			0.48			&			0.1			&			0.07			&			1.66			&			-0.11			&			C	\\
2941				&			ADS520 	 			&			9.57			&			0.22			&			0.7			&			0.7			&			1.87			&			1.87			&			28.23			&			0.00			&			A	\\
5842				&			HD 7693  			&			23.4			&			0.04			&			0.89			&			0.84			&			5.96			&			5.81			&			57.89			&			-0.07			&			D	\\
7078				&			HD 9021				&			0.64			&			0.31			&			1.21			&			0.7			&			0.12			&			0.09			&			1.97			&			-0.09			&			C	\\
7751				&			HD 10360 			&			52.2			&			0.53			&			0.77			&			0.75			&			5.61			&			5.54			&			173.15			&			-0.54			&			D	\\
7918				&			HD 10307			&			7.1			&			0.42			&			0.8			&			0.14			&			1.26			&			0.57			&			22.13			&			-2.75			&			E	\\
8903				&			HD 11636			&			0.66			&			0.9			&			2.07			&			1.28			&			0.01			&			0.01			&			2.37			&			-0.18			&			E	\\
11231				&			HD 15064			&			0.64			&			0.29			&			1.01			&			0.68			&			0.12			&			0.1			&			1.95			&			-0.06			&			C	\\
12062				&			HD 15862			&			2.04			&			0.26			&			0.95			&			0.44			&			0.43			&			0.3			&			6.11			&			-0.32			&			C	\\
12153				&			HD 16234			&			4.22			&			0.88			&			11			&			9.41			&			0.09			&			0.08			&			15.11			&			-0.39			&			E	\\
12623				&			HD 16739			&			1.27			&			0.66			&			1.39			&			1.13			&			0.1			&			0.09			&			4.35			&			-0.12			&			E	\\
\bf{14954}			&			\bf{HD 19994}			&			120			&			0.26			&			1.35			&			0.35			&			27.34			&			14.83			&			354.86			&			-28.27			&			F	\\
18512				&			HD 24916			&			174.55			&			0			&			0.35			&			0.17			&			52.36			&			37.68			&			315.65			&			0.00			&			G	\\
20087				&			HD 27176			&			7.05			&			0.17			&			1.76			&			0.95			&			1.66			&			1.26			&			20.08			&			-0.65			&			E	\\
20935				&			HD 28394			&			0.99			&			0.24			&			1.13			&			1.11			&			0.19			&			0.19			&			2.95			&			0.00			&			C	\\
24419				&			HD 34101			&			1.75			&			0.08			&			0.9			&			0.21			&			0.52			&			0.27			&			4.52			&			-0.17			&			C	\\
30920		&				&		4.3		&		0.37		&		0.22		&		0.08		&		0.773		&		0.488		&		13.421		&		-1.1144		&	K\\
33451				&			HD 51825			&			9.3			&			0.43			&			1.61			&			1.26			&			1.31			&			1.17			&			29.93			&			-0.75			&			E	\\
34164				&			HD 53424			&			1.7			&			0.27			&			1.09			&			0.66			&			0.34			&			0.27			&			5.13			&			-0.19			&			C	\\
39893				&							&			1.81			&			0.21			&			0.95			&			0.52			&			0.4			&			0.31			&			5.29			&			-0.19			&			C	\\
44248		&				&		10.4		&		0.15		&		1.44		&		0.89		&		2.470		&		1.983		&		29.246		&		-0.6940		&	K\\
45343		&				&		97.2		&		0.28		&		0.52		&		0.51		&		17.405		&		17.251		&		295.520		&		-0.4459		&	K\\
56809				&			HD 101177			&			240.39			&			0.05			&			1.95			&			1.36			&			63.9			&			54.21			&			605.53			&			-5.06			&			G	\\
63406				&			HD 112914			&			1.59			&			0.33			&			0.82			&			0.23			&			0.32			&			0.18			&			4.86			&			-0.44			&			C	\\
64241		&				&		11.8		&		0.5		&		1.3		&		1.12		&		1.398		&		1.306		&		38.807		&		-0.6585		&	K\\
64797		&				&		89.2		&		0.12		&		0.73		&		0.52		&		21.553		&		18.458		&		245.009		&		-3.5692		&	K\\
\bf{67275}			&			\bf{HD 120136}			&			245			&			0.91			&			1.35			&			0.4			&			4.38			&			2.52			&			868.91			&			-147.85			&			F	\\
67422		&				&		32.7		&		0.45		&		0.72		&		0.65		&		4.306		&		4.109		&		105.959		&		-1.1531		&	K\\
72848				&			HD 131511   			&			0.52			&			0.51			&			0.93			&			0.45			&			0.07			&			0.05			&			1.71			&			-0.13			&			D	\\
73440				&			HD 133621			&			1.25			&			0.22			&			1.03			&			0.15			&			0.32			&			0.14			&			3.56			&			-0.30			&			C	\\
75379				&			HD 137502			&			0.91			&			0.68			&			1.26			&			0.68			&			0.07			&			0.05			&			3.12			&			-0.26			&			C	\\
76852				&			HD 140159			&			12.4			&			0.15			&			2			&			1.98			&			2.7			&			2.69			&			34.96			&			-0.02			&			E	\\
79101				&			HD 145389			&			2.24			&			0.47			&			3.47			&			1.31			&			0.33			&			0.21			&			7.23			&			-0.68			&			C	\\
80346				&							&			2.07			&			0.67			&			0.5			&			0.13			&			0.18			&			0.1			&			6.98			&			-1.04			&			C	\\
80686				&			HD 147584			&			0.12			&			0.06			&			1.05			&			0.37			&			0.04			&			0.02			&			0.3			&			-0.01			&			C	\\
82817				&			HD 152771			&			1.38			&			0.05			&			0.56			&			0.33			&			0.38			&			0.3			&			3.47			&			-0.04			&			E	\\
82860				&			HD 153597			&			0.33			&			0.21			&			1.18			&			0.52			&			0.08			&			0.05			&			0.96			&			-0.05			&			C	\\
84425		&				&		7.7		&		0.49		&		1.23		&		0.86		&		0.970		&		0.824		&		25.221		&		-0.9998		&	K\\
84720				&			HD 156274			&			91.65			&			0.78			&			0.79			&			0.47			&			4.33			&			3.41			&			321.18			&			-24.55			&			D	\\
84949				&			HD 157482			&			4.87			&			0.67			&			2.62			&			1.15			&			0.29			&			0.42			&			16.6			&			-1.73			&			E	\\
86400				&			HD 1360346			&			0.39			&			0.23			&			0.72			&			0.39			&			0.08			&			0.06			&			1.15			&			-0.05			&			C	\\
86722				&			HD 161198			&			3.97			&			0.94			&			0.94			&			0.34			&			0.04			&			0.03			&			14.21			&			-2.18			&			E	\\
87895				&			HD 163840			&			2.14			&			0.41			&			0.99			&			0.68			&			0.32			&			0.27			&			6.84			&			-0.25			&			C	\\
91768				&			HD 173739			&			49.51			&			0.53			&			0.39			&			0.34			&			5.43			&			5.1			&			164.2			&			-2.67			&			G	\\
\bf{93017}			&			\bf{ADS 11871}			&			22.96			&			0.25			&			1.65			&			1.58			&			4.34			&			4.26			&			68.77			&			-0.21			&			A	\\
93825		&				&		32.7		&		0.32		&		1.27		&		1.25		&		5.463		&		5.424		&		101.163		&		-0.1364		&	K\\
95028				&			HD 181602			&			0.85			&			0.37			&			1.4			&			0.5			&			0.15			&			0.1			&			2.65			&			-0.22			&			C	\\
95575				&			HD 183255			&			0.62			&			0.15			&			0.78			&			0.38			&			0.15			&			0.11			&			1.74			&			-0.06			&			C	\\
\bf{98001}			&			\bf{HD 188753}			&			11.65			&			0.47			&			1.3			&			1.11			&			1.48			&			1.38			&			37.98			&			-0.66			&			E	\\
99965				&			HD 193216			&			1.24			&			0.08			&			0.88			&			0.56			&			0.32			&			0.26			&			3.26			&			-0.05			&			C	\\
109176				&			HD 210027			&			0.12			&			0			&			1.25			&			0.8			&			0.03			&			0.03			&			0.22			&			0.00			&			C	\\
111170				&			HD 213429			&			1.74			&			0.38			&			1.08			&			0.7			&			0.28			&			0.23			&			5.5			&			-0.22			&			C	\\
113718				&			HD 217580			&			1.16			&			0.54			&			0.76			&			0.18			&			0.15			&			0.08			&			3.78			&			-0.51			&			C	\\
\bf{116310}			&			\bf{HD 221673}			&			95			&			0.322			&			2			&			2			&			15.7			&			15.7			&			294.14			&			0.00			&			H	\\
\bf{116727}			&			\bf{HD 222404}			&			18.5			&			0.36			&			1.59			&			0.4			&			3.55			&			1.9			&			57.04			&			-5.72			&			F	\\
\bf{}				&			\bf{Kepler 16}			&			0.22			&			0.16			&			0.69			&			0.2			&			0.06			&			0.03			&			0.63			&			-0.03			&			I	\\
\bf{}				&			\bf{Kepler 34}			&			0.23			&			0.52		&			1.05			&			1.02			&			0.03			&			0.03			&			0.76			&			-0.003			&			J	\\
		&		Kepler 35		&		0.18		&		0.14		&		0.81		&		0.81		&		0.040		&		0.040		&		0.504		&		0.0000		&	L\\
		&		Kepler 38		&		0.15		&		0.1		&		0.95		&		0.25		&		0.043		&		0.024		&		0.397		&		-0.0163		&	L\\
		&		Kepler 47		&		0.08		&		0.02		&		1.043		&		0.362		&		0.025		&		0.015		&		0.185		&		-0.0021		&	L\\
		&		Kepler 64		&		0.17		&		0.21		&		1.528		&		0.408		&		0.042		&		0.023		&		0.490		&		-0.0333		&	L\\

\hline
\end{tabular}
\caption{ Orbital parameters and stability radii. Columns are: (1) HIP name, (2) Alternative name (some cases), 
(3) Semi-major axis, (4) eccentricity, (5) Mass of the primary star, (6) Mass of the secondary star, 
(7) Outer stable circumstellar radius for primary star, (8) Outer stable radius for secondary star, 
(9) Inner stable radius for circumbinary zone and (10) References about the orbital parameters. 
References are: A$=$\protect\cite{HW99}, B$=$\protect\cite{CI09}, C$=$\protect\cite{JJ05}, 
D$=$\protect\cite{BD07}, E$=$\protect\cite{Martin98}, F$=$\protect\cite{DB07}, G$=$\protect\cite{SL04}, 
H$=$\protect\cite{MF10}, I$=$\protect\cite{DC11}, J$=$\protect\cite{LINES11}, K$=$\protect\cite{HS13} 
and L$=$\protect\cite{Eggl13}  }
\end{table*}

\begin{table*}
\begin{tabular}{ccccccccccccccc}
\label{tabla-todo}

HIP    &    Alter. Name    &    case    &    $R_{inner}^{*1}$    &    $R_{outer}^{*1}$    &    $R_{inner}^{*2}$    &    $R_{outer}^{*2}$    &    $R_{inner}^{**}$    &    $R_{outer}^{**}$\\
    &        &        &    $AU$    &    $AU$    &    $AU$    &    $AU$    &    $AU$    &    $AU$\\
\hline

--            	&	            V821 Cas        	&	3	&	-	&	-	&	-	&	-	&	5.04	&	        8.55    \\
1955	&	            HD 2070            	&	3	&	-	&	-	&	-	&	-	&	1.77	&	     2.20 \\
2941	&	            ADS520             	&	1	&	0.55	&	0.95	&	0.55	&	0.95	&	-	&	        -    \\
5842	&	            HD 7693         	&	1	&	0.87	&	1.5	&	0.78	&	1.34	&	-	&	        -    \\
7078	&	            HD 9021            	&	3	&	-	&	-	&	-	&	-	&	2.06	&	     2.67 \\
7751	&	            HD 10360        	&	1	&	0.66	&	1.13	&	0.63	&	1.08	&	-	&	        -    \\
7918	&	            HD 10307        	&	1	&	0.71	&	1.22	&	0.03	&	0.04	&	-	&	        -    \\
8903	&	            HD 11636        	&	3	&	-	&	-	&	-	&	-	&	5.01	&	     7.66 \\
12153	&	            HD 16234        	&	3	&	-	&	-	&	-	&	-	&	128.82	&	        219.85    \\
\bf{14954}        	&	            \bf{HD 19994}        	&	1	&	1.92	&	3.3	&	0.15	&	0.25	&	-	&	        -    \\
18512	&	            HD 24916        	&	1	&	0.15	&	0.25	&	0.04	&	0.06	&	-	&	        -    \\
30920		&				&	1	&	0.06	&	0.1	&	0.01	&	0.02	&	-	&	-\\
44248		&				&	1	&	2.19	&	2.47	&	0.9	&	1.55	&	-	&	-\\
45343		&				&	1	&	0.31	&	0.54	&	0.3	&	0.52	&	-	&	-\\
56809	&	            HD 101177        	&	1	&	3.86	&	6.63	&	1.95	&	3.34	&	-	&	        -    \\
64797		&				&	1	&	0.6	&	1.03	&	0.31	&	0.54	&	-	&	-\\
67422		&				&	1	&	0.58	&	1	&	0.48	&	0.82	&	-	&	-\\
\bf{67275}        	&	            \bf{HD 120136}        	&	1	&	1.92	&	3.3	&	0.19	&	0.33	&	-	&	        -    \\
80346	&	                        	&	1	&	-	&	-	&	0.03	&	0.04	&	-	&	        -    \\
80686	&	            HD 147584        	&	3	&	-	&	-	&	-	&	-	&	1.23	&	        1.90    \\
82817	&	            HD 152771        	&	1	&	0.36	&	0.38	&	0.14	&	0.24	&	-	&	        -    \\
82860	&	            HD 153597        	&	3	&	-	&	-	&	-	&	-	&	1.63	&	        2.42    \\
84720	&	            HD 156274        	&	1	&	0.69	&	1.19	&	0.26	&	0.44	&	-	&	        -    \\
91768	&	            HD 173739        	&	1	&	0.18	&	0.31	&	0.14	&	0.24	&	-	&	        -    \\
\bf{93017}        	&	            \bf{ADS 11871}        	&	1	&	2.86	&	4.34	&	2.64	&	4.26	&	-	&	        -    \\
93825		&				&	1	&	1.72	&	2.94	&	1.66	&	2.85	&	-	&	-\\
95028	&	            HD 181602        	&	3	&	-	&	-	&	-	&	-	&	2.87	&	     3.25 \\
109176	&	            HD 210027        	&	3	&	-	&	-	&	-	&	-	&	1.83	&	        2.99    \\
\bf{116310}        	&	            \bf{HD 221673}        	&	1	&	4.06	&	6.97	&	4.06	&	6.97	&	-	&	        -    \\
\bf{116727}        	&	            \bf{HD 222404}        	&	1	&	2.62	&	4.5	&	0.2	&	0.32	&	-	&	        -    \\
\bf{}            	&	            \bf{Kepler 16}        	&	3	&	-	&	-	&	-	&	-	&	0.66	&	        0.85    \\
\bf{}				&			\bf{Kepler 34}			&	3	&	-	&	-	&	-	&	-	&	1.67	&	2.81\\
		&		Kepler 35		&	3	&	.	&	-	&	-	&	-	&	2.44	&	2.75\\
		&		Kepler 38		&	3	&	.	&	-	&	-	&	-	&	0.71	&	0.97\\
		&		Kepler 47		&	3	&	.	&	-	&	-	&	-	&	0.185	&	0.19\\
		&		Kepler 64		&	3	&	.	&	-	&	-	&	-	&	2.6	&	4.41\\

\hline
\end{tabular}
\caption{Binary systems with efective habitable zone. In these cases we can 
find radiative safe zones (at apo and periastron) within the stable zone. Columns are:
  (1) HIP name, (2) Alternative name (some cases), (3) Case of HZ
  introduced in section \ref{HZ}, (4) Maximum of inner radii at
  Apoastron and Periastron for the primary star intersected with the
  corresponding stability zone, (5) Minimum of outer radii at
  Apoastron and Periastron for the primary star intersected with the
  corresponding stability zone, (6) Maximum of inner radii at
  Apoastron and Periastron for the secondary star intersected with the
  corresponding stability zone, (7) Minimum of outer radii at
  Apoastron and Periastron for the secondary star intersected with the
  corresponding stability zone, (8) Maximum of inner radii at
  Apoastron and Periastron for the circumbinary zone intersected with
  the corresponding stability zone, (9) Minimum of outer radii at
  Apoastron and Periastron for the circumbinary zone intersected with
  the corresponding stability zone. Corrections because of the shift
  were taken in to account. }
\end{table*}

%---------------------------------------------------------------------------------------------------------------------------
%
%\acknowledgments 
%
We thank the anonymous referee for a very throrough review of our
manuscript and suggestions that resulted in a clearer and deeper
exposition. We acknowledge financial support from UNAM/DGAPA through
grant IN114114.

%---------------------------------------------------------------------------------------------------------------------------

\end{document}